\documentclass[a4paper,fleqn]{cas-sc}

\usepackage[numbers]{natbib}
\usepackage{graphicx} 
\usepackage{braket}
\usepackage{amsmath}
\usepackage{amssymb}
\usepackage{mathtools}
\usepackage{booktabs}
\usepackage{bm}
\usepackage{xurl}
\usepackage{subcaption}
\usepackage{listings}
\def\tsc#1{\csdef{#1}{\textsc{\lowercase{#1}}\xspace}}
\tsc{WGM}
\tsc{QE}
\tsc{EP}
\tsc{PMS}
\tsc{BEC}
\tsc{DE}

\begin{document}
\let\WriteBookmarks\relax
\def\floatpagepagefraction{1}
\def\textpagefraction{.001}

\shorttitle{Fast memory-efficient classical simulation of QML}

\shortauthors{Y.~Kawase}

\title [mode = title]{Fast and memory-efficient classical simulation of quantum machine learning via forward and backward gate fusion}




%
\author[1]{Yoshiaki Kawase}[type=editor,
                        orcid=0000-0002-4471-0851]

\cormark[1]

\fnmark[1]

\ead{ykawase@g.ecc.u-tokyo.ac.jp}



\affiliation[1]{organization={Graduate School of Information Science and Technology, The University of Tokyo},
    addressline={7-3-1 Hongo}, 
    city={Bunkyo-ku, Tokyo},
    postcode={113-8656}, 
    country={Japan}}

\cortext[cor1]{Corresponding author}

\begin{abstract} 
While real quantum devices have been increasingly used to conduct research focused on achieving quantum advantage or quantum utility in recent years, 
executing deep quantum circuits or performing quantum machine learning with large-scale data on current noisy intermediate-scale quantum devices remains challenging, 
making classical simulation essential for quantum machine learning research. 
However, such classical simulation often suffers from the cost of gradient calculations, requiring enormous memory or computational time.
To address these problems, we propose a method to fuse multiple consecutive gates in each of the forward and backward paths to improve throughput by minimizing global memory accesses. 
As a result, we achieved approximately $20$ times throughput improvement for a Hardware-Efficient Ansatz with $12$ or more qubits, reaching over $30$ times improvement on a mid-range consumer GPU with limited memory bandwidth. 
By combining our proposed method with gradient checkpointing, 
we drastically reduced memory usage, making it possible to train a large-scale quantum machine learning model, a $20$-qubit, $1{,}000$-layer model with $60{,}000$ parameters, using $1{,}000$ samples in approximately $20$ minutes per epoch. 
This implies that we can train the model on large datasets, comprising tens of thousands of samples, like MNIST or CIFAR-10, within a realistic time frame (e.g., $20$ hours per epoch). 
Thus, our proposed method significantly accelerates such classical simulations, making a significant contribution to advancing research in quantum machine learning and variational quantum algorithms, such as verifying algorithms on large datasets or investigating learning theories of deep quantum circuits like barren plateaus. 
\end{abstract}

\begin{keywords}
classical simulation \sep quantum machine learning \sep 
variational quantum algorithms \sep GPU \sep Triton \sep kernel fusion
\end{keywords}

\maketitle

\section{Introduction}\label{sec:intro}
In recent years, quantum machine learning (QML), such as quantum neural networks \cite{farhi2018classification, mitarai2018quantum}, and variational quantum algorithms (VQAs) \cite{cerezo2021variational}, such as variational quantum eigen-solver (VQE) \cite{peruzzo2014variational, kandala2017hardware} and quantum approximate optimization algorithm (QAOA) \cite{farhi2014quantum}, have been actively researched. 
Current noisy intermediate-scale quantum devices have limitations in the depth of quantum circuits and the connectivity of qubits,
so classical simulation is still important to verify new algorithms or investigate learning theories, including barren plateaus \cite{mcclean2018barren, wang2021noise, cerezo2021cost} and the existence of many local minima \cite{you2021exponentially, bittel2021training}, particularly in deep quantum circuits. 

Most of the current research on classical simulation of quantum circuits \cite{pednault2019leveraging, tindall2024efficient} focuses on simulating a single large quantum circuit, supporting the achievement of quantum advantage \cite{arute2019quantum, google2025observation} or quantum utility \cite{kim2023evidence}. 
In general, a classical simulation using a state vector \cite{suzuki2021qulacs, qiskit2024} requires exponentially large memory with respect to the number of qubits, 
so it is usual to utilize a supercomputer or multi-GPU clusters \cite{pednault2019leveraging, de2019massively, kim2025scaleqsim, haner20175, smelyanskiy2016qhipster}, 
or to use storage devices \cite{pednault2019leveraging, park2022snuqs} for classical simulation over $30$ qubits. 
The existing classical simulators are superior at computing a single large quantum circuit, 
but classical simulation of QML requires memory that grows linearly with batch size, as well as exponentially with the number of qubits, 
making it challenging even for a $20$-qubit quantum circuit. 
Moreover, there is a challenge in calculating gradients \cite{mitarai2018quantum, jones2020efficient}, as it requires enormous computation time or memory. 
However, most of the existing general-purpose quantum circuit simulators are limited to optimizing a single quantum circuit, or are not specifically designed for batch processing, often failing to take advantage of the parallel performance of GPUs due to inefficient memory access. 

A popular approach for gradient calculation is the parameter-shift rule \cite{mitarai2018quantum}. 
This approach is memory-efficient, but the number of entire quantum circuit executions scales linearly with the number of variational gates, making it impractical for large-scale models in classical simulation. 
On the other hand, the adjoint method is fast because the effective number of entire quantum circuit executions is independent of the number of variational gates. 
However, a practical implementation of the adjoint method faces a critical trade-off between memory usage and execution time regarding the management of intermediate states. 
A naive implementation stores all intermediate states in the forward path in order to reuse them in the backward path, 
causing memory usage to increase linearly with the number of gates. 
Alternatively, a memory-efficient implementation avoids storing intermediate states by sequentially applying each gate to recompute them during the backward path  \cite{jones2020efficient, luo2020yao}, leading to a significant increase in computational time due to the repeated accesses to exponentially large quantum states \cite{jones2020efficient, luo2020yao}. 
Therefore, existing simulators typically optimize only forward paths using gate fusion \cite{suzuki2021qulacs, haner20175, smelyanskiy2016qhipster}, or sequentially apply quantum gates to obtain intermediate states in the backward paths \cite{jones2020efficient, luo2020yao}, leaving the backward path bottlenecked by inefficient memory accesses.

In this paper, we propose a method to fuse multiple consecutive quantum gates into a single fused operator for each of the forward and backward paths, without writing any intermediate states of the fused gates to global memory. 
This approach allows us to reduce the memory requirements for storing quantum states and decrease the number of accesses to global memory, leading to an enhancement of the arithmetic intensity. 
However, if we simply read the input state and temporarily store all recomputed intermediate states, the number of variables to be held in registers can easily exceed the register capacity.  
This causes register spilling and increases the computation time. 
Therefore, to reduce this register pressure, we minimize the number of intermediate states held in registers, and load unitary matrices whenever necessary, overwriting the previous values to reuse registers. 
As a result, our proposed method achieves approximately $20$ times improvement in throughput compared to a PyTorch native implementation for a typical single-layer Hardware-Efficient Ansatz (HEA) \cite{kandala2017hardware}. 
In particular, for a mid-range GPU with limited memory bandwidth, we achieved over $30$ times improvement in throughput, 
highlighting the effectiveness of our proposed method on GPUs with limited memory bandwidth. 

Moreover, since our implementation uses Triton \cite{triton2019} and PyTorch \cite{Ansel_PyTorch_2_Faster_2024} and is integrated with PyTorch's automatic differentiation, 
it is easy to utilize the existing PyTorch ecosystem. 
For example, by combining our proposed method with gradient checkpointing, we can reduce memory usage to $O(\sqrt{d})$ for $d$ HEA layers. 
In our numerical experiment, we performed a task of minimizing a sum of batched expectation values for an observable at the final states of a $20$-qubit with $1{,}000$ HEA layers ($60{,}000$ parameters). 
As a result, we succeeded in training the model on $1{,}000$ samples in approximately $20$ minutes per epoch. 
This result indicates that we can train a model consisting of $60{,}000$ parameters using tens of thousands of samples within $20$ hours per epoch, i.e., a realistic time frame. 
Further acceleration of training can be easily realized using multiple GPUs with PyTorch's distributed training utilities, such as DistributedDataParallel. 

In this way, since our proposed method enables the training of QML with large datasets or deep quantum circuits within a realistic timeframe, it is expected to significantly accelerate research and development in QML.

\section{Preliminaries}
In this section, we explain the adjoint method for efficiently calculating gradients. 

\subsection{State vector simulation and objective function}\label{subsec:state_vec_sim}
Here, we consider a quantum state of an $n$-qubit system represented by a complex vector of length $2^n$, $|\psi\rangle \in \mathbb{C}^{2^n}$, 
and a parameterized quantum circuit (PQC) consisting of a series of $M$ parameterized unitary gates $U(\boldsymbol{\theta}) = U_M(\theta_M) \cdots U_1(\theta_1)$ characterized by trainable parameters $\boldsymbol{\theta} = \{\theta_1, \dots, \theta_M\}$. 
We represent the initial state as $|\psi_0 \rangle$, 
the intermediate state as $|\psi_j \rangle \coloneqq U_j \ldots U_1 |\psi_0\rangle$, 
and the final state as $|\psi_M (\boldsymbol{\theta})\rangle \coloneqq U(\boldsymbol{\theta})|\psi_0\rangle$. 

As is often the case in QML and VQAs, we focus on the task of minimizing a loss function $L$ by optimizing the trainable parameters $\boldsymbol{\theta}$. 
Let $O$ denote an observable. 
We represent the expectation value $E(\boldsymbol{\theta})$ as follows: 
\begin{equation}
   E(\boldsymbol{\theta}) = \langle \psi_M(\boldsymbol{\theta}) | O | \psi_M(\boldsymbol{\theta}) \rangle.
\end{equation}
Although the loss function $L$ is a function of $E(\boldsymbol{\theta})$ in general, we assume $L=E$ in our benchmarks for simplicity. 

\subsection{Adjoint method}\label{sec:adjoint_method}
As we mentioned in Sec.~\ref{sec:intro}, the adjoint method can efficiently calculate gradients in classical simulation. 
In this subsection, we explain the adjoint method for optimizing trainable parameters $\boldsymbol{\theta}$ in a PQC, $U(\boldsymbol{\theta})$. 
Specifically, we explain how to calculate gradients for the $j$-th unitary gate $U_j(\theta_j)$. 

In the adjoint method, we introduce an adjoint state $|\lambda_j\rangle$ for the backward path. 
For the $j$-th unitary gate $U_j(\theta_j)$, the adjoint state is propagated by 
\begin{equation}\label{eqn:update_adjoint_state}
|\lambda_{j-1}\rangle = U_j^\dagger (\theta_j) |\lambda_j\rangle. 
\end{equation}
Using this adjoint state $|\lambda_j\rangle$, the gradient of the loss function $L$ with respect to $\theta_j$ can be described by 
\begin{equation}\label{eq:grad_theta}
    \frac{\partial L}{\partial \theta_j} = \text{Re} \left\langle \lambda_j \left| \frac{\partial U_j}{\partial \theta_j} \right| \psi_{j-1} \right\rangle.
\end{equation}

Note that from Eq.~\eqref{eq:grad_theta}, we need both the input quantum state $|\psi_{j-1}\rangle$ and the adjoint state $|\lambda_j\rangle$ 
to calculate the gradient $\frac{\partial L}{\partial \theta_j}$. 
This indicates that we need to preserve the input quantum states $|\psi_{j-1}\rangle$ to efficiently compute the gradients. 
Namely, a naive implementation requires $O(M \cdot 2^n)$ memory to store all the input quantum states $\{|\psi_0\rangle, \dots, |\psi_{M-1}\rangle\}$ in memory. 
As the number of variational parameters $M$ grows, the required memory can easily exceed the memory capacity, making classical simulation infeasible. 
Even within the memory capacity, the computation time significantly increases due to repeated accesses to these exponentially large vectors in global memory.   
We will tackle these problems with our proposed method in the following sections. 

Also note that, since we assume $L=E$, the adjoint state at the final gate can be described by 
\begin{equation}\label{eqn:bwd_exp_val}
    |\lambda_M\rangle = 2 O |\psi_{M}\rangle . 
\end{equation}


\section{Methods}
In this section, we describe the implementation of classical simulation of a quantum circuit, focusing specifically on the efficient usage of GPU memory. 
Since the performance of classically simulating a quantum circuit is typically limited by the memory bandwidth, 
we explain a kernel fusion technique to reduce the data transfer, and a recomputation technique to prevent register spilling. 

\subsection{Applying a quantum gate}\label{subsec:gate_basics}
As we mentioned in Sec.~\ref{subsec:state_vec_sim}, 
an $n$-qubit state vector $|\psi\rangle$ is represented by a complex vector of length $2^n$. 
We describe $|\psi\rangle$ using computational basis states $|x\rangle$ and complex probability amplitudes $a_x$, whose index $x$ $(0\leq x<2^n)$ corresponds to its binary representation $x_{n-1}\ldots x_0$ $(x=\sum_{i=0}^{n-1} x_i 2^i)$ of length $n$, as follows: 
\begin{equation*}
    |\psi\rangle = \sum_{x=0}^{2^n-1} a_x |x\rangle = a_{0\ldots00} |0\cdots00\rangle + a_{0\ldots01} |0\cdots01\rangle + \cdots 
    + a_{1\ldots10} |1\cdots10\rangle + a_{1\ldots11} |1\cdots11\rangle
\end{equation*}

Next, we consider applying a single qubit gate $u$ on the $t$-th qubit, where $u$ is represented by
\begin{equation*}
    u = 
    \begin{pmatrix}
    u_{00} & u_{01} \\
    u_{10} & u_{11}
    \end{pmatrix}, 
\end{equation*}
and $U$ is represented by $U=I\otimes\cdots\otimes \underbrace{u}_{t\text{-th}} \otimes\cdots\otimes I$. 

This operation can be described by a linear transformation for pairs of probability amplitudes ($a_{\ldots0_t\ldots}$, $a_{\ldots1_t\ldots}$), 
whose indexes only differ at the $t$-th bit, updating them to ($a'_{\ldots0_t\ldots}$, $a'_{\ldots1_t\ldots}$) as follows: 
\begin{equation*}
    \begin{pmatrix}
    a'_{...0_t...} \\
    a'_{...1_t...} 
    \end{pmatrix}
    =
    \begin{pmatrix}
    u_{00} & u_{01} \\
    u_{10} & u_{11}
    \end{pmatrix}
    \begin{pmatrix}
    a_{...0_t...} \\
    a_{...1_t...} 
    \end{pmatrix}, 
\end{equation*}
where the bits in the dotted parts of the binary representation remain unchanged. 
In a parallelized implementation on a GPU, each thread typically loads a pair of amplitudes from global memory, applies the unitary matrix, and writes the resulting amplitudes back to global memory. 
This operation is performed for all of the $2^{n-1}$ pairs, requiring a huge number of global memory accesses. 

Next, we explain how to update amplitudes when applying two-qubit gates. 
First, when applying a {\it CZ} gate acting on control qubit $c$ and target qubit $t$ to a quantum state, 
we only need to multiply the amplitudes $a_{\ldots 1_c \ldots 1_t \ldots}$ (or $a_{\ldots 1_t \ldots 1_c \ldots}$) by $-1$, 
where the control and target bits of the index are both $1$. 
Similarly, when applying {\it CNOT} gate acting on control qubit $c$ and target qubit $t$ to a quantum state, 
we can update the quantum state by swapping the pair of amplitudes $a_{\ldots 1_c \ldots 0_t \ldots}$ and $a_{\ldots 1_c \ldots 1_t \ldots}$ (or $a_{\ldots 0_t \ldots 1_c \ldots}$ and $a_{\ldots 1_t \ldots 1_c \ldots}$), i.e., swapping the amplitudes corresponding to the $t$-th bit being $0$ and $1$ for all indexes whrere the $c$-th bit is $1$. 

Therefore, whenever we apply a quantum gate, we must read and write $O(2^n)$ probability amplitudes, which limits the performance of classical simulation on GPUs due to memory bandwidth constraints.

\subsection{Gate fusion in the forward path}
To save memory bandwidth in the forward path, 
a popular technique is to fuse multiple quantum gates before applying them to a quantum state \cite{suzuki2021qulacs, haner20175, smelyanskiy2016qhipster}. 
In this paper, to maintain consistency with the calculation in the backward path, 
we consider applying $m$ consecutive single-qubit gates, each denoted as $u^{(1)},\ldots,u^{(m)}$, acting on the same target qubit $t$, such that: 
\begin{align*}
    \begin{pmatrix}
    a'_{...0_t...} \\
    a'_{...1_t...} 
    \end{pmatrix}
    &= u^{(m)} \cdots u^{(1)} 
    \begin{pmatrix}
    a_{...0_t...} \\
    a_{...1_t...} 
    \end{pmatrix} \\
    &=
    \underbrace{
    \begin{pmatrix}
    u_{00}^{(m)} & u_{01}^{(m)} \\
    u_{10}^{(m)} & u_{11}^{(m)}
    \end{pmatrix}
    \cdots
    \begin{pmatrix}
    u_{00}^{(1)} & u_{01}^{(1)} \\
    u_{10}^{(1)} & u_{11}^{(1)}
    \end{pmatrix}
    }_{\text{Fused Unitary}}
    \begin{pmatrix}
    a_{...0_t...} \\
    a_{...1_t...} 
    \end{pmatrix} .
\end{align*}
By performing this operation, we can update a quantum state by reading and writing its state vector only once per fused gate consisting of $m$ consecutive single-qubit gates.
This operation allows us to reduce memory accesses by a factor of $m$. 
To accelerate the classical simulation of typical Ansatzes in QML and VQA, in this paper, 
we consider applying a fused gate consisting of consecutive single-qubit gates acting on the same qubit and also the adjacent qubits. 

Although the adjoint method requires the input state vector for each variational gate, 
our implementation stores only the output state vector of each fused variational gate during the forward path. 
As detailed in the next subsection, we then use this stored output state vector to recompute the input state vector for each constituent gate within the fused operator. 

When storing these state vectors, we can store their real and imaginary parts with original precision (\texttt{float32} or \texttt{float64}), 
or with reduced precision (\texttt{bfloat16} or \texttt{float32}) to further reduce the memory usage and use larger batch sizes. 
We refer to this reduced-precision approach as the \textit{memory-saving mode} (denoted as Triton fused (mem save) in our numerical experiments).
Even when we store them with halved precision, 
we perform the calculations in original high precision using (\texttt{float32} or \texttt{float64}), i.e., the precision is only reduced when storing them in global memory. 
This technique allows us to mitigate the impact on computational accuracy while halving memory usage.
Furthermore, in our numerical experiments, we will demonstrate the efficient training of deep quantum circuits by reducing the storage of state vectors using our proposed method combined with the gradient checkpointing technique. 

\subsection{Gate fusion in the backward path}
Next, we explain the gate fusion technique for the backward path. 
One of the major contributions of our work is the realization of efficient gate fusion in the backward path. 

As shown in Eq.~\eqref{eq:grad_theta}, 
when we calculate the gradients with respect to the variational parameters $\theta_{j+k}$, 
we need both the input quantum state $|\psi_{j+k-1}\rangle$ and the adjoint state $|\lambda_{j+k}\rangle$ for the corresponding variational gate. 
If we simply fuse $m$ consecutive variational gates in the backward path, 
it requires storing the elements of quantum states $|\psi_j\rangle,\ldots,|\psi_{j+m-1}\rangle$, $|\lambda_j\rangle,\ldots,|\lambda_{j+m}\rangle$ in registers. 
Therefore, as $m$ increases, 
the required memory can easily exceed the register capacity of the GPU, leading to register spilling into a low-speed memory region, which increases computation time. 

\begin{figure}
    \centering
    \includegraphics[width=0.85\linewidth]{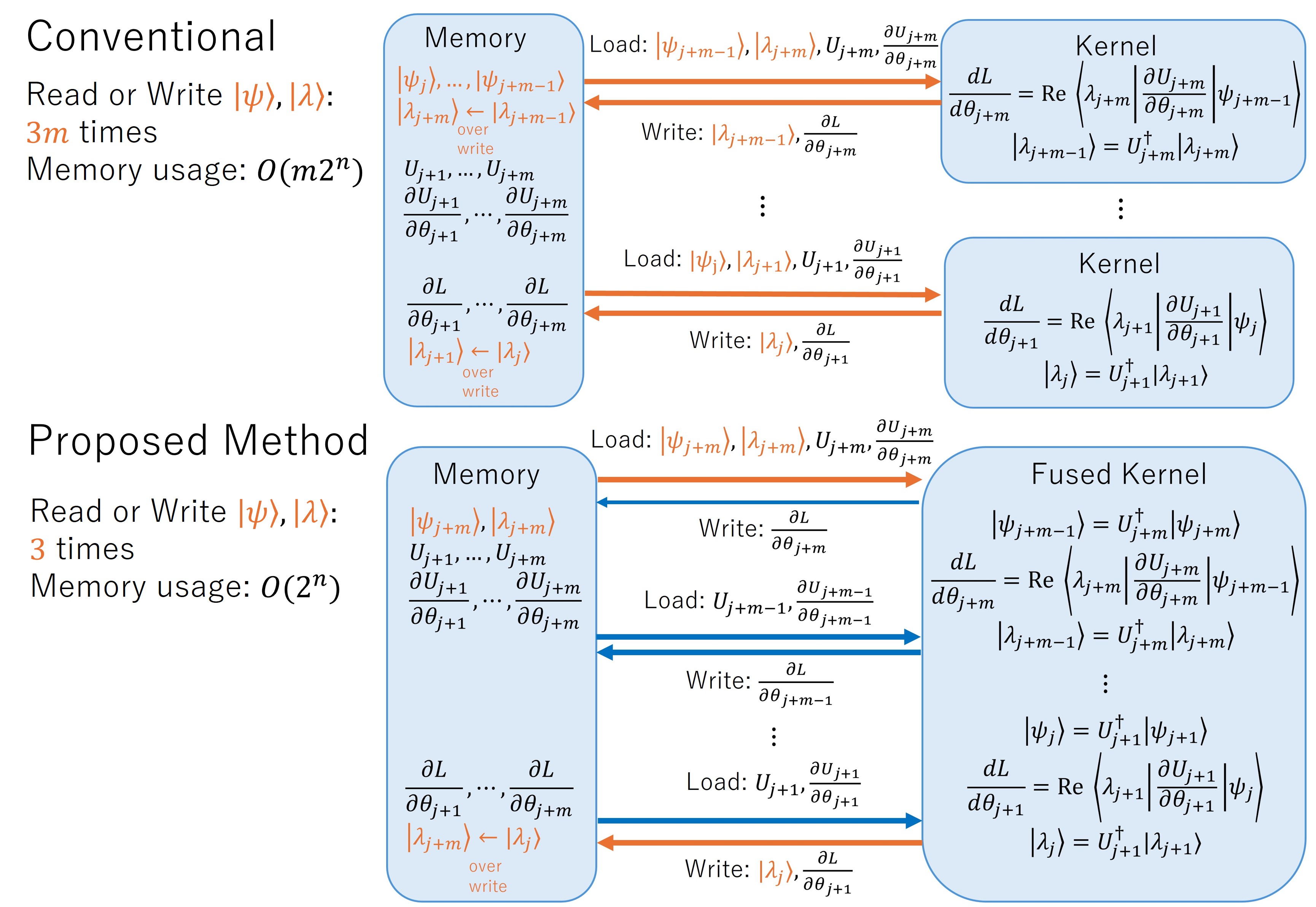}
    \caption{Conventional method and proposed method}
    \label{fig:method}
\end{figure}

To address this issue, by focusing on the reversibility of quantum gates, 
we recompute the intermediate states within GPU registers when needed, instead of storing all of them in global memory. 
Specifically, as shown in Fig.~\ref{fig:method}, 
given the quantum state $|\psi_{j+m}\rangle$ (stored during the forward path) and the adjoint state $|\lambda_{j+m}\rangle$ 
as inputs for the fused gates consisting of $m$ consecutive variational gates, 
each thread initially loads its assigned pairs of complex elements, 
$\left(a^{(j+m)}_{\dots 0_t \dots}, a^{(j+m)}_{\dots 1_t \dots}\right)$ and $\left(r^{(j+m)}_{\dots 0_t \dots}, r^{(j+m)}_{\dots 1_t \dots}\right)$, 
from global memory into its local registers, 
where $a^{(j+m)}_x$ represents the probability amplitude of the $(j+m)$-th quantum state $|\psi_{j+m} \rangle$ at index $x$
and $r^{(j+m)}_x$ represents the complex element of the $(j+m)$-th adjoint state $|\lambda_{j+m} \rangle$ at index $x$. 
Then, within a single fused GPU kernel, each thread performs the following procedure without accessing global memory to obtain intermediate states.

To recompute the input quantum states for $U_{j+k}(\theta_{j+k})$ and calculate the gradients $\frac{\partial L}{\partial \theta_{j+k}}$ for $k=1,\ldots,m$, 
we iterate the gate index of the fused gates in reverse order from $k=m$ to $k=1$. 
For each step $k$:  
\begin{enumerate}
\item \textbf{Updating quantum states:}\\
    Each thread loads the elements of the unitary matrix $u_{j+k}(\theta_{j+k})$ and applies the matrix to its assigned pair of amplitudes of $|\psi_{j+k}\rangle$ in order to calculate 
    the corresponding amplitudes of $ |\psi_{j+k-1} \rangle= U_{j+k}^\dagger (\theta_{j+k}) |\psi_{j+k}\rangle $: 
    \begin{equation*}
        \begin{pmatrix}
        a^{(j+k-1)}_{...0_t...} \\
        a^{(j+k-1)}_{...1_t...} 
        \end{pmatrix} 
        = u_{j+k}^\dagger (\theta_{j+k})         
        \begin{pmatrix}
        a^{(j+k)}_{...0_t...} \\
        a^{(j+k)}_{...1_t...} 
        \end{pmatrix} , 
    \end{equation*}
    where $a^{(j+k)}_x$ represents the probability amplitude of the $(j+k)$-th intermediate state $|\psi_{j+k} \rangle$ at index $x$. 
    
\item \textbf{Calculating gradients:}\\
    For a variational single-qubit rotation gate, 
    we can easily calculate $\frac{\partial u_{j+k}(\theta_{j+k})}{\partial \theta_{j+k}}=-\frac{1}{2}\sin{\frac{\theta_{j+k}}{2}}I-i \frac{1}{2} \cos{\frac{\theta_{j+k}}{2}} P$ 
    from the definition $u_{j+k}(\theta_{j+k})=\cos{\frac{\theta_{j+k}}{2}}I-i\sin{\frac{\theta_{j+k}}{2}} P$, where $P \in \{I,X,Y,Z\}$. \\
    According to Eq.~\eqref{eq:grad_theta}, 
    each thread calculates its local contribution to the gradient $\frac{\partial L}{\partial \theta_{j+k}}$: 
    \begin{equation*}
        \text{Re} \left[
        \begin{pmatrix}
        r^{(j+k)*}_{...0_t...} & r^{(j+k)*}_{...1_t...} 
        \end{pmatrix} 
        \frac{\partial u_{j+k}(\theta_{j+k})}{\partial \theta_{j+k}}
        \begin{pmatrix}
        a^{(j+k-1)}_{...0_t...} \\
        a^{(j+k-1)}_{...1_t...} 
        \end{pmatrix} 
        \right], 
    \end{equation*}
    where $r^{(j+k)}_x$ represents the complex element of the $(j+k)$-th intermediate adjoint state $|\lambda_{j+k} \rangle$ at index $x$ and $^*$ denotes the complex conjugate. 
\item \textbf{Updating adjoint state:}\\
According to $|\lambda_{j+k-1} \rangle= U_{j+k}^\dagger (\theta_{j+k}) |\lambda_{j+k}\rangle$, 
each thread updates its assigned pair of components of the adjoint state $|\lambda_{j+k}\rangle$ using the adjoint of the loaded gate $u_{j+k}(\theta_{j+k})$:
\begin{equation*}
    \begin{pmatrix}
    r^{(j+k-1)}_{...0_t...} \\
    r^{(j+k-1)}_{...1_t...} 
    \end{pmatrix} 
    = u_{j+k}^\dagger (\theta_{j+k}) 
    \begin{pmatrix}
    r^{(j+k)}_{...0_t...} \\
    r^{(j+k)}_{...1_t...} 
    \end{pmatrix} . 
\end{equation*}
\end{enumerate}
By repeating this procedure, we can efficiently calculate all the gradients of the training parameters $\boldsymbol{\theta}$. 

Using this procedure, 
each thread only needs to hold a pair of probability amplitudes of the intermediate state $|\psi_{j+k-1}\rangle$ 
and a pair of complex elements of the adjoint state $|\lambda_{j+k}\rangle$, 
along with the elements of the unitary matrix $u_{j+k}(\theta_{j+k})$ and its derivative, in registers during kernel processing. 
To avoid register spilling, we load the elements of the unitary matrix $u_{j+k}(\theta_{j+k})$ and its derivative only when required, 
and reuse the registers by overwriting the previous elements of the state vector, the adjoint state, the unitary matrix, and its derivative. 
Compared to the method of storing all the input quantum states and adjoint states, 
although our method requires additional calculations to recompute the input quantum states for each $U_{j+k}(\theta_{j+k})$, 
it allows us to mitigate the bottleneck, i.e., minimize the accesses to global memory. 
This is expected to reduce the required memory and improve throughput, as fewer quantum states need to be stored.

\subsection{Theoretical memory analysis with gradient checkpointing}\label{subsec:theoretical_mem}
By combining our proposed method with gradient checkpointing (specifically using PyTorch's \path{torch.utils.checkpoint}), 
we can significantly reduce the memory usage, especially for deep quantum circuits.

To theoretically analyze the behavior of memory usage, we consider a typical deep parameterized quantum circuit consisting of a repeating layer structure, 
such as a Hardware-Efficient Ansatz (HEA). 
Let $d$ be the total number of HEA layers in a quantum circuit, and let $l_\text{var}$ and $l_\text{const}$ represent the number of variational gates and constant gates per HEA layer, respectively. 

In the PyTorch native implementation of the adjoint method without gradient checkpointing, 
since we need to store the input quantum states of all gates for the backward path, 
the number of required state vectors to be stored is $(l_\text{var}+l_\text{const})d$. 
To mitigate this linear memory growth, we apply gradient checkpointing by dividing the quantum circuit into checkpoint blocks, each consisting of $b$ consecutive HEA layers. 
For simplicity, we assume that $d$ is divisible by $b$. 

When combining the PyTorch native implementation with gradient checkpointing, 
the number of required state vectors to be stored for the backward path becomes $(l_\text{var}+l_\text{const})b+d/b$. 
The first term represents the stored intermediate states within a single active checkpoint block, 
and the second term ($d/b$) arises from the stored input quantum states for each checkpoint block, which are required to recalculate the intermediate state vectors within the checkpoint block during gradient checkpointing. 
Therefore, letting $M_\text{sv}$ be the required memory for a single state vector, 
the required memory $M_{total}$ (bytes) for the PyTorch native implementation is estimated as 
\begin{align}\label{eqn:mem_cap_pytorch}
    M_{total}(b) &\simeq \left ( (l_\text{var}+l_\text{const})  \cdot b + \frac{d}{b} \right) \times M_\text{sv} \\
    &\geq 2\sqrt{(l_\text{var}+l_\text{const})d} \times M_\text{sv} , 
\end{align}
where the equality holds when $b^*=\sqrt{\frac{d}{l_\text{var}+l_\text{const}}}$. 

\begin{figure}
    \centering
    \includegraphics[width=0.85\linewidth]{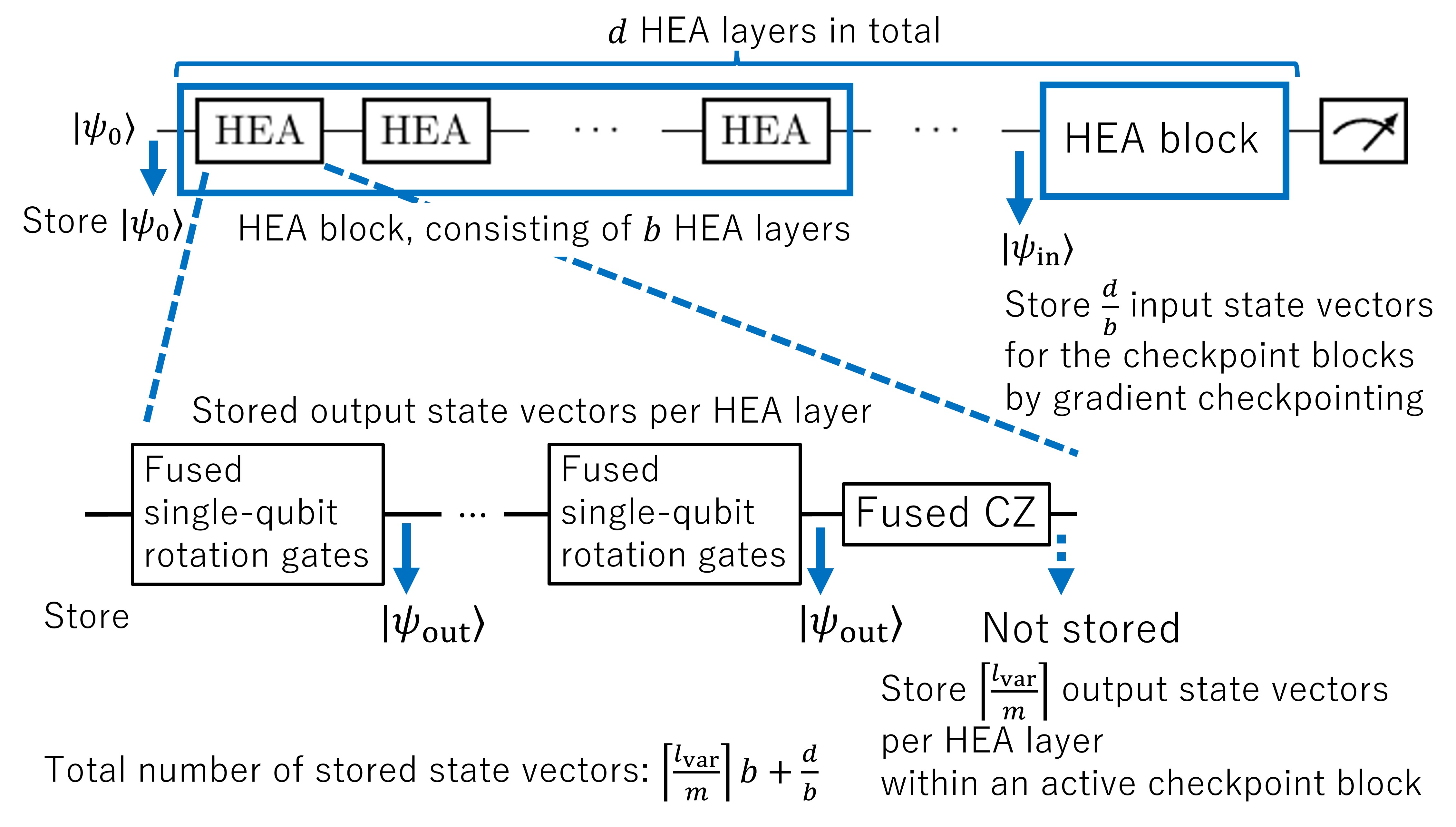}
    \caption{Stored state vectors when combining our proposed method with gradient checkpointing}
    \label{fig:checkpoint}
\end{figure}

Similarly, as illustrated in Fig.~\ref{fig:checkpoint}, for our proposed method combined with gradient checkpointing, 
we must store $d/b$ input quantum states for the checkpoint blocks. 
Additionally, within the active checkpoint block, we only need to store $\left \lceil \frac{l_\text{var}}{m} \right \rceil \cdot b$ output state vectors of the fused variational gates, 
since the implementation of our proposed method does not require storing any input and output state vectors for constant gates. 
By introducing $\alpha$ as an effective number of fused gates to uniformly treat the memory reduction effects regardless of the type of saved tensors, 
the required memory (bytes) can be estimated as 
\begin{align}\label{eqn:mem_cap_proposed}
    M_{total}(b) &\simeq \left (\left \lceil \frac{l_\text{var}}{\alpha} \right \rceil \cdot b + \frac{d}{b} \right) \times M_\text{sv} \\
    &\geq 2\sqrt{\left \lceil \frac{l_\text{var}}{\alpha} \right \rceil d} \times M_\text{sv}, 
\end{align}
where the equality holds when $b^*=\sqrt{d/\left \lceil \frac{l_\text{var}}{\alpha} \right \rceil }$.  
Specifically, for our proposed method (Triton fused), $\alpha$ corresponds to $m$. 
For our proposed method with memory-saving mode (Triton fused (mem save)), we can effectively regard $\alpha=2m$ 
because we store the real and imaginary parts of the state vector as \texttt{bfloat16} for \texttt{complex64} or \texttt{float32} for \texttt{complex128} to halve the memory required to store state vectors. 
Note that this memory-saving technique applies only to the stored output state vectors of each fused variational gate within an active checkpoint block. 
The input quantum states for checkpoint blocks are stored without type conversion because they are automatically managed by PyTorch's checkpointing mechanism. 

Consequently, by combining our proposed method with gradient checkpointing, the memory usage scales as $O(\sqrt{d})$. 
This combined approach allows us to drastically reduce the total memory requirements while maintaining the high throughput achieved by minimizing global memory accesses, enabling efficient classical simulation of deep quantum circuits. 

\section{Numerical experiments}
In this section, to evaluate the performance of our proposed method, implemented with Triton \cite{triton2019} and PyTorch \cite{Ansel_PyTorch_2_Faster_2024}, 
we measured the average execution time across multiple runs over a $1{,}000$ ms window using the benchmarking utility \path{triton.testing.do_bench} in Triton \cite{triton2019}. 
If a single execution exceeded this duration, the single execution time was evaluated. 
We also measured peak memory usage using \path{torch.cuda.max_memory_allocated} from a single measurement,  
since peak memory usage remains consistent across multiple executions. 
Note that although our implementation supports both single precision (\texttt{float32}/\texttt{complex64}) and double precision (\texttt{float64}/\texttt{complex128}), 
we use \texttt{float32} (\texttt{complex64}) in our numerical experiments to focus on performance evaluation for practical QML tasks, which often have severe memory constraints. 
Furthermore, throughout our numerical experiments, we apply quantum circuits to randomly initialized quantum states $|\psi_0\rangle$ rather than $|0\rangle^{\otimes n}$ 
in order to avoid artificial speedups caused by sparsity and to reflect the actual computational load and memory bandwidth utilization. 

\subsection{The effect of gate fusion}
\begin{figure}
    \centering
    \includegraphics[width=0.9\linewidth]{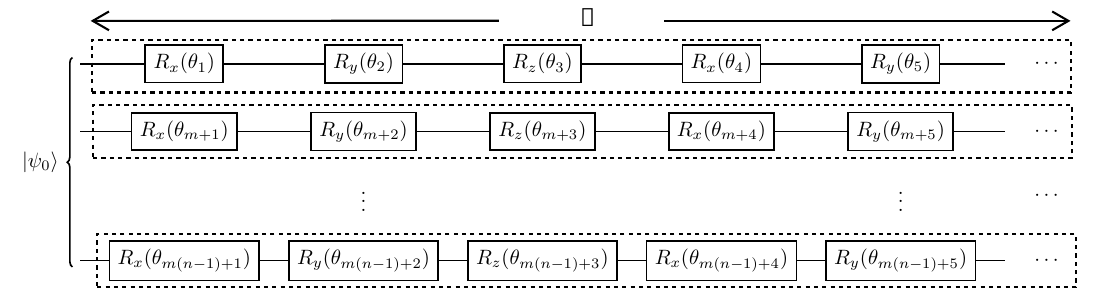}
    \caption{A quantum circuit consisting of $m$ consecutive {\it Rx}, {\it Ry}, and {\it Rz} gates used for the benchmark of verifying the effect of our proposed method}
    \label{fig:n_single_qubit_gates}
\end{figure}

In this subsection, 
we investigate the effect of the gate fusion on a quantum circuit consisting of consecutive single-qubit gates, as shown in Fig.~\ref{fig:n_single_qubit_gates}. 
Using a $20$-qubit system and a batch size of $100$, we focus on how memory usage and execution time depend on the number of gates to be fused in the forward and backward paths.
In the numerical experiments in this subsection, 
while the PyTorch native implementation sequentially applies $m$ consecutive single-qubit gates acting on each qubit in the repeating order of  {\it Rx}, {\it Ry}, and {\it Rz}, 
our proposed method fuses these $m$ gates into a single operator. 

\begin{figure}[h!]
    \centering
    \begin{subfigure}[b]{0.47\textwidth}
    \centering
    \includegraphics[width=0.95\linewidth]{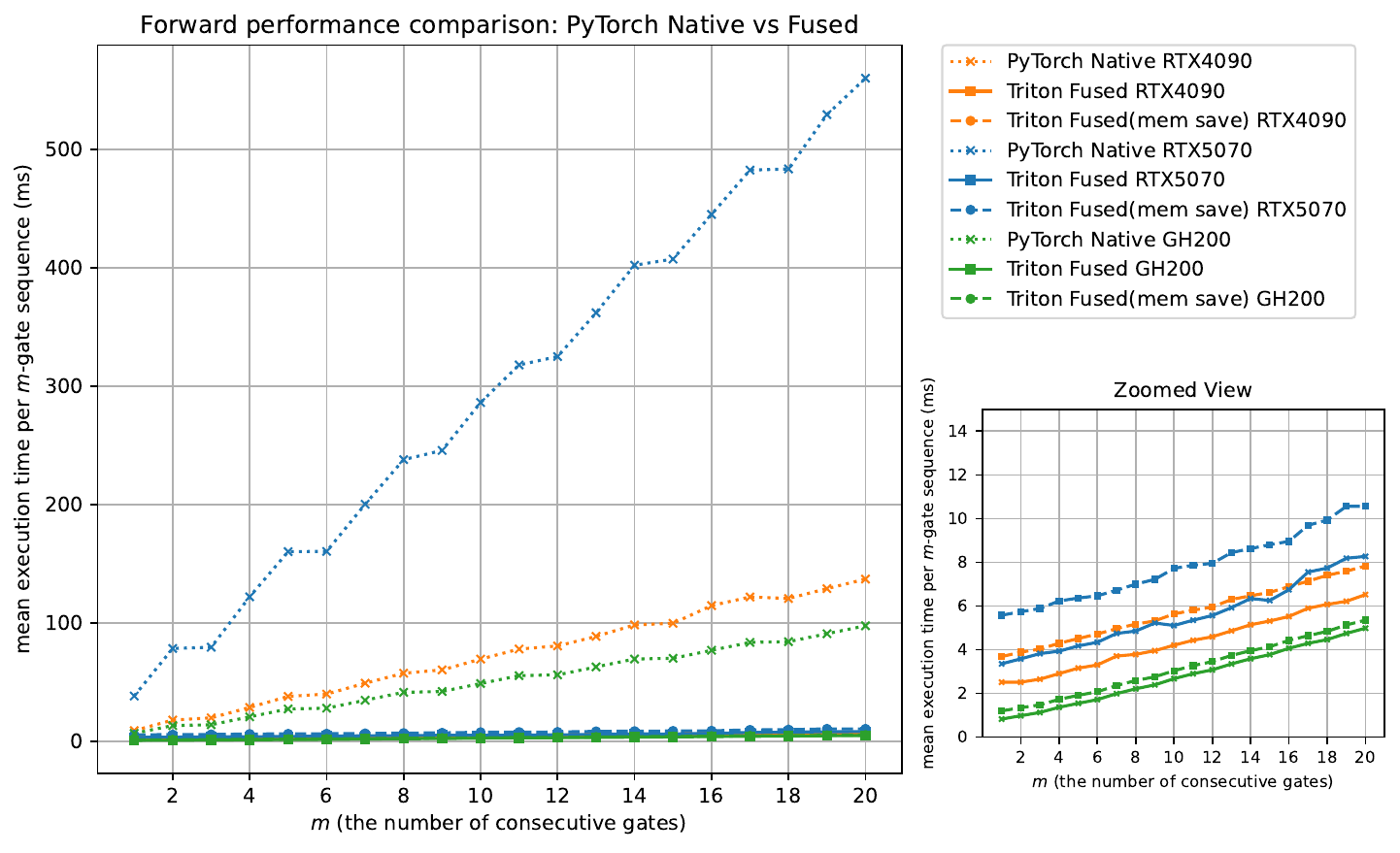}
    \caption{Execution time per $m$-gate sequence}
    \label{fig:fwd_time}
    \end{subfigure}
    \hfill %
    \begin{subfigure}[b]{0.47\textwidth}
    \centering
    \includegraphics[width=0.95\linewidth]{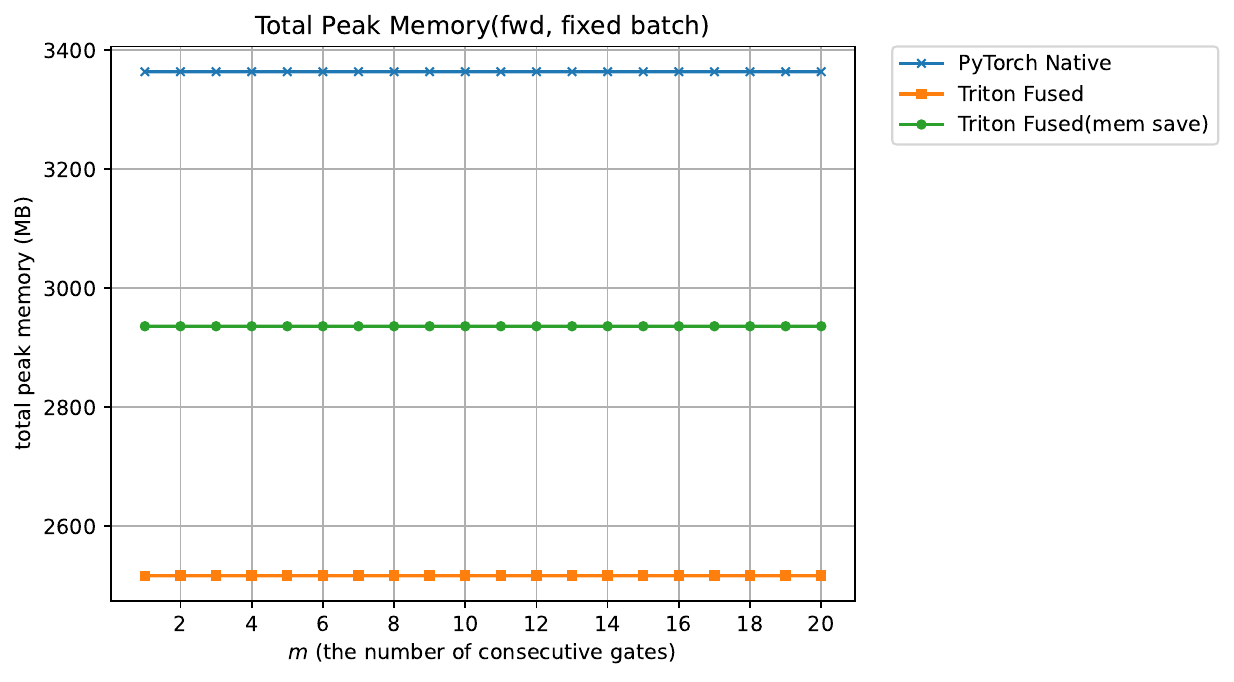}
    \caption{Total peak memory usage}
    \label{fig:fwd_mem}
    \end{subfigure}
    \caption{ (a) Execution time per $m$-gate sequencee and (b) Total peak memory usage when applying $m$ consecutive single-qubit gates on each qubit during the forward path }
\end{figure}

First, we measured the mean execution time to apply these $m$ gates during the forward path. 
As shown in Fig.~\ref{fig:fwd_time}, 
the execution time of our proposed method increases almost linearly with $m$, 
but much more gradually than that of the PyTorch native implementation.
Consequently, our proposed method achieves a drastic acceleration. 
For the PyTorch native implementation, 
we observe that the execution time increases significantly when applying the {\it Rx} and {\it Ry} gates, 
whereas it increases gradually only when applying the diagonal matrix {\it Rz} gate.

The total peak memory usage after applying the $m$-gate sequence to all $20$ qubits during the forward path is shown in Fig.~\ref{fig:fwd_mem}. 
As shown, across all implementations, the peak memory usage remains almost constant regardless of the number of consecutive single-qubit gates to be applied. 
In the forward path, the peak memory usage of our proposed method (Triton fused) is the lowest among the three implementations. 
The peak memory usage is $2{,}500$ MB, roughly corresponding to three batched state vectors ($3\times 800$ MB). 
It appears that the input, output, and temporary batched state vectors are held, and the remaining $100$ MB is internal overhead. 
Similarly, the Triton fused (mem\_save) implementation used an additional $400$ MB, 
caused by storing the batched state vector converted to \texttt{bfloat16} for the backward path. 
The PyTorch native implementation consumes approximately $800$ MB more memory than the Triton fused implementation, corresponding to a single batched state vector. 
This suggests that the PyTorch native implementation temporarily keeps a copy of the batched state vectors during the forward path,  
likely due to explicitly cloning them for safety. 
Since PyTorch's automatic differentiation requires out-of-place operations on inputs to preserve the computation graph, 
our implementation is considered to be memory-efficient.

\begin{figure}[h!]
    \centering
    \begin{subfigure}[b]{0.45\textwidth}
        \centering
        \includegraphics[width=0.95\linewidth]{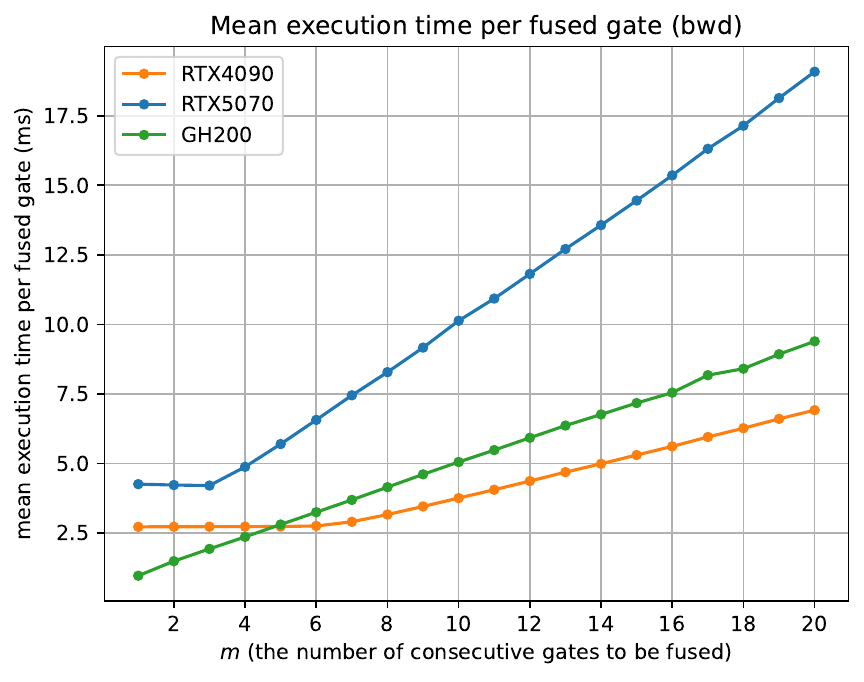}
        \caption{Execution time per fused gate}
        \label{fig:backward_exec_time}
    \end{subfigure}
    \hfill %
    \begin{subfigure}[b]{0.45\textwidth}
        \centering
        \includegraphics[width=0.8\linewidth]{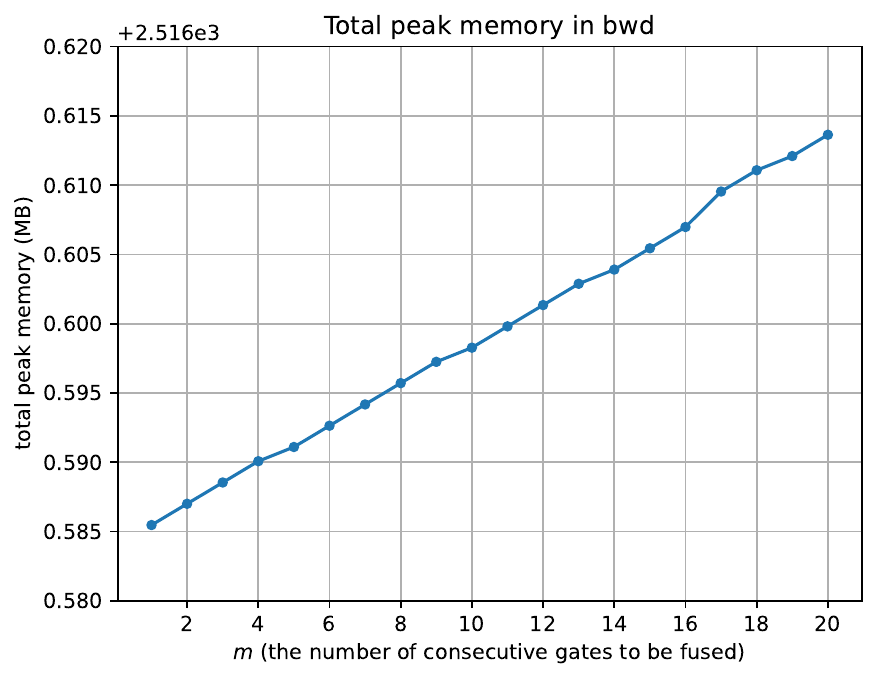}
        \caption{Total peak memory usage}
        \label{fig:backward_mem_eff}
    \end{subfigure}
    \caption{ (a) Execution time per fused gate and (b) Total peak memory usage when applying $m$ consecutive single-qubit gates on each qubit at the backward path }
\end{figure}

Next, to evaluate the performance of the backward function itself, 
we apply the same quantum circuit as above, shown in Fig.~\ref{fig:n_single_qubit_gates}, 
and measure the mean execution time per fused backward kernel. 
From the results shown in Fig.~\ref{fig:backward_exec_time}, 
for consumer GPUs such as the RTX 5070 and RTX 4090, 
the computation time of the backward calculation for a fused gate consisting of a few constituent gates is almost constant, i.e., the recomputation time is negligible 
because the memory access latency to read quantum states and adjoint states from global memory is dominant. 
As the number of gates to be fused further increases, the computation time begins to scale linearly. 
On the other hand, for data center GPUs with large memory bandwidth such as the GH200, 
the computation time increases linearly with the number of gates to be fused from the beginning. 
Fig.~\ref{fig:backward_mem_eff} shows the total peak memory usage after applying the functions to all $20$ qubits. 
As shown, the peak memory usage scales linearly but very gradually with the number of gates to be fused, growing by only $30$ KB (from $2516.585$ MB to $2516.613$ MB) even when $m$ is increased to $20$. 
This indicates that the computation of the fused gates in the backward path is processed efficiently while mitigating the use of temporary variables. 

\begin{figure}[h!]
    \centering
    \begin{subfigure}[b]{0.3\textwidth}
        \centering
        \includegraphics[width=0.95\linewidth]{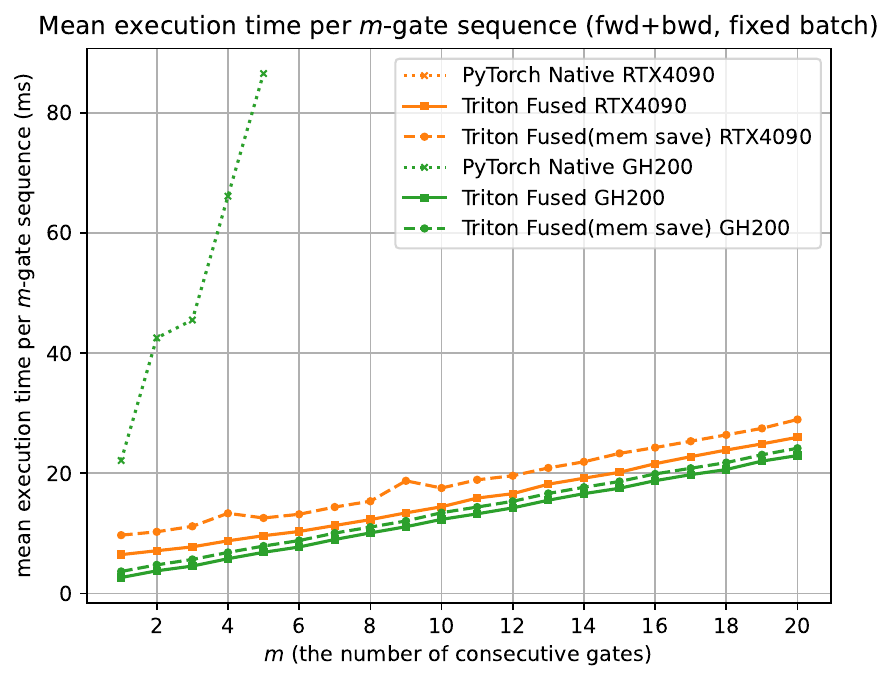}
        \caption{Execution time per $m$-gate sequence}
        \label{fig:fwdbwd_time}
    \end{subfigure}
    \hfill %
    \begin{subfigure}[b]{0.3\textwidth}
        \centering
        \includegraphics[width=0.95\linewidth]{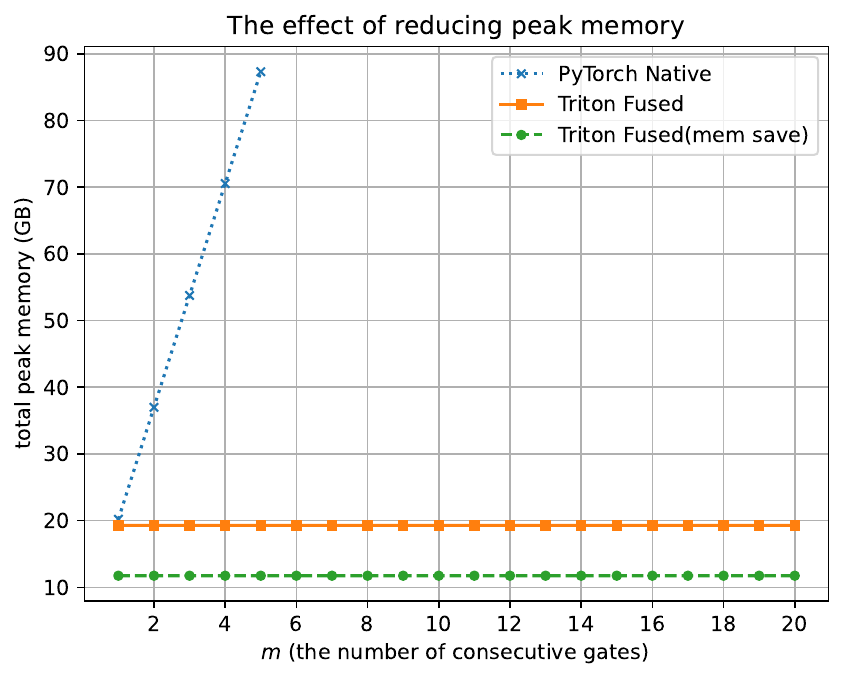}
        \caption{Total peak memory usage}
        \label{fig:fwdbwd_mem}
    \end{subfigure}
    \hfill %
    \begin{subfigure}[b]{0.3\textwidth}
        \centering
        \includegraphics[width=0.95\linewidth]{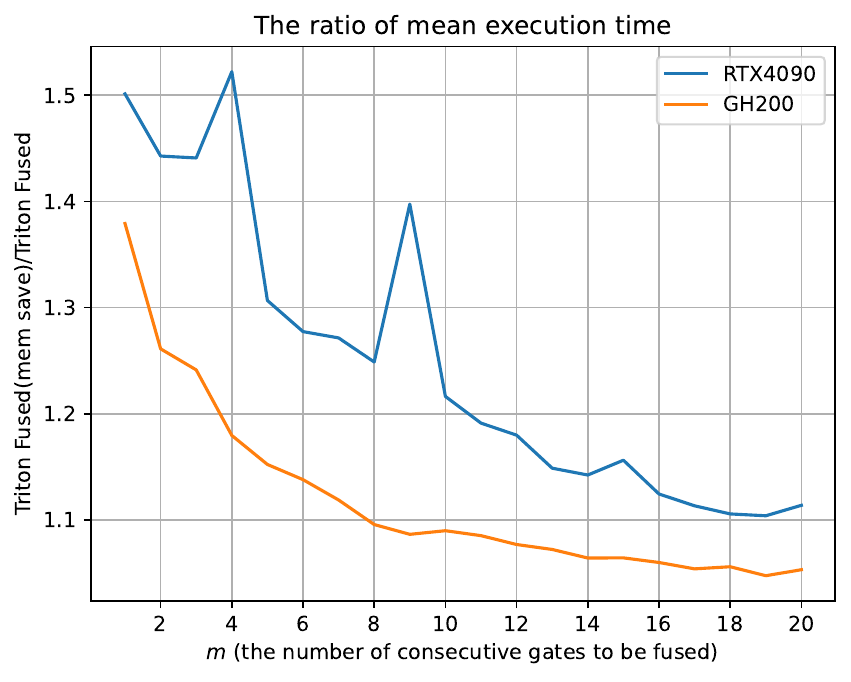}
        \caption{Execution time ratio}
        \label{fig:fwdbwd_time_ratio}
    \end{subfigure}
    \caption{ (a) Execution time per $m$-gate sequence and (b) total peak memory usage (c) execution time ratio between Triton Fused and Triton Fused (mem save) 
    when minimizing the sum of expectation values at the final quantum state after applying the fused single-qubit gates shown in Fig.~\ref{fig:n_single_qubit_gates} to a $100$ fixed batch of random initial states for one epoch. 
    Note that, unlike the smoother trend of the data center GPU GH200, local performance fluctuations (e.g., at $m=4,9$) are observed on the RTX 4090, likely due to background OS tasks and dynamic frequency scaling. 
    }
\end{figure}

Finally, we measured the mean execution time per $m$-gate sequence and total peak memory usage over the forward and backward paths 
while performing a task of minimizing a sum of expectation values for the observable $O=IXYZ \ldots IXYZ$ (repeated $5$ times) 
using the same quantum circuit as above, shown in Fig.~\ref{fig:n_single_qubit_gates}, with $20$ qubits and a batch size of $100$.  
The mean execution time per $m$-gate sequence is shown in Fig.~\ref{fig:fwdbwd_time}, 
and the total peak memory usage after applying the consecutive gates to all $20$ qubits is shown in Fig.~\ref{fig:fwdbwd_mem}. 
We performed the evaluation of the PyTorch native implementation for this numerical experiment using only GH200 
because it requires a huge amount of memory to store the state vectors. 
Also, we evaluated our proposed method on the RTX 4090 and GH200 due to the insufficient memory capacity on the RTX 5070.

As shown in Fig.~\ref{fig:fwdbwd_time}, 
the implementation of our proposed method performs significantly faster than the PyTorch native implementation. 
The mean execution time per $m$-gate sequence for both approaches increases linearly with the number of consecutive variational gates, but that for our proposed method increases more gradually. 
This trend suggests that reducing the number of accesses to global memory by our proposed method contributes to this acceleration. 
Additionally, for the result of PyTorch native implementation, the increase in execution time is gradual when an {\it Rz} gate is added. 

As shown in Fig.~\ref{fig:fwdbwd_mem}, 
the peak memory usage of the PyTorch native implementation shows a steep linear increase with the number of consecutive gates, 
since all input quantum states for the gates are stored during the forward path for later use in the backward path. 
For our proposed method, on the other hand, 
the required memory remains almost constant regardless of the number of variational gates to be fused, and is drastically lower than that of the PyTorch native implementation, 
since our proposed method only needs to store the resulting state of the fused variational gates. 

In addition, the Triton fused (mem save) implementation further reduced the peak memory usage  
by type converting the state vector from \texttt{float32} to \texttt{bfloat16} when temporarily storing it in the forward path. 
As shown in Fig.~\ref{fig:fwdbwd_time_ratio}, when $10$ or more variational gates are fused, 
the mean execution time per $m$-gate sequence of Triton fused (mem save) is $5\%$ to $10\%$ slower than that of Triton fused implementation on the GH200, 
and $10\%$ to $20\%$ slower on the RTX 4090. 
Importantly, we can reduce peak memory usage by $40\%$ at the cost of an approximate $10\%$ increase in computation time. 
This trade-off is practically important, particularly for a consumer GPU with limited memory capacity and bandwidth. 

Note that in Fig.~\ref{fig:fwdbwd_time} and Fig.~\ref{fig:fwdbwd_time_ratio}, 
local fluctuations (e.g., at $m=4,9$) are observed on the consumer GPU RTX 4090, 
whereas the data center GPU GH200 exhibits a smoother trend. 
These performance fluctuations are typical of consumer GPUs and are caused by background OS interrupts, display rendering tasks, and dynamic voltage and frequency scaling (DVFS), which occasionally compete with GPU resources and memory bandwidth.

\subsection{Throughput for a Hardware Efficient Ansatz}
\begin{figure}
    \centering
    \includegraphics[width=0.85\linewidth]{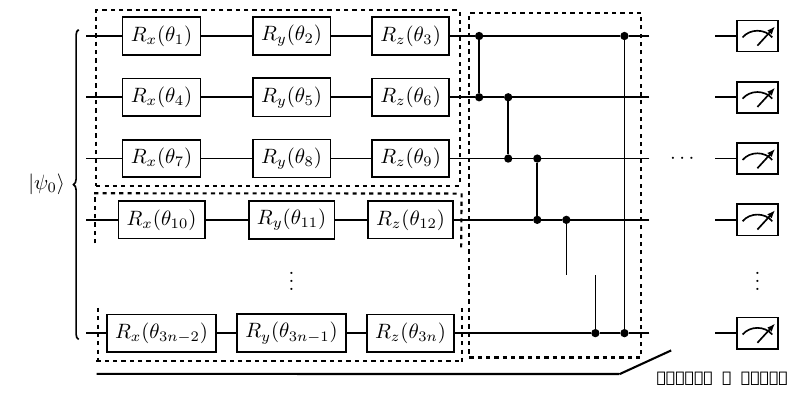}
    \caption{HEA and gate fusion method: 
    the quantum circuit consists of single-qubit gates ({\it Rx}, {\it Ry}, {\it Rz}) and {\it CZ} gates. 
    The dotted frames indicate the range of the gate fusion in our proposed method. 
    We fused up to nine single-qubit gates, including single-qubit gates acting on the same and adjacent qubits, 
    and also the consecutive {\it CZ} gates in the implementation of our proposed method. 
    In our numerical experiments, we used a quantum circuit consisting of $d$ layers of this HEA. 
    }
    \label{fig:hea}
\end{figure}

\begin{figure}
    \centering
    \includegraphics[width=0.9\linewidth]{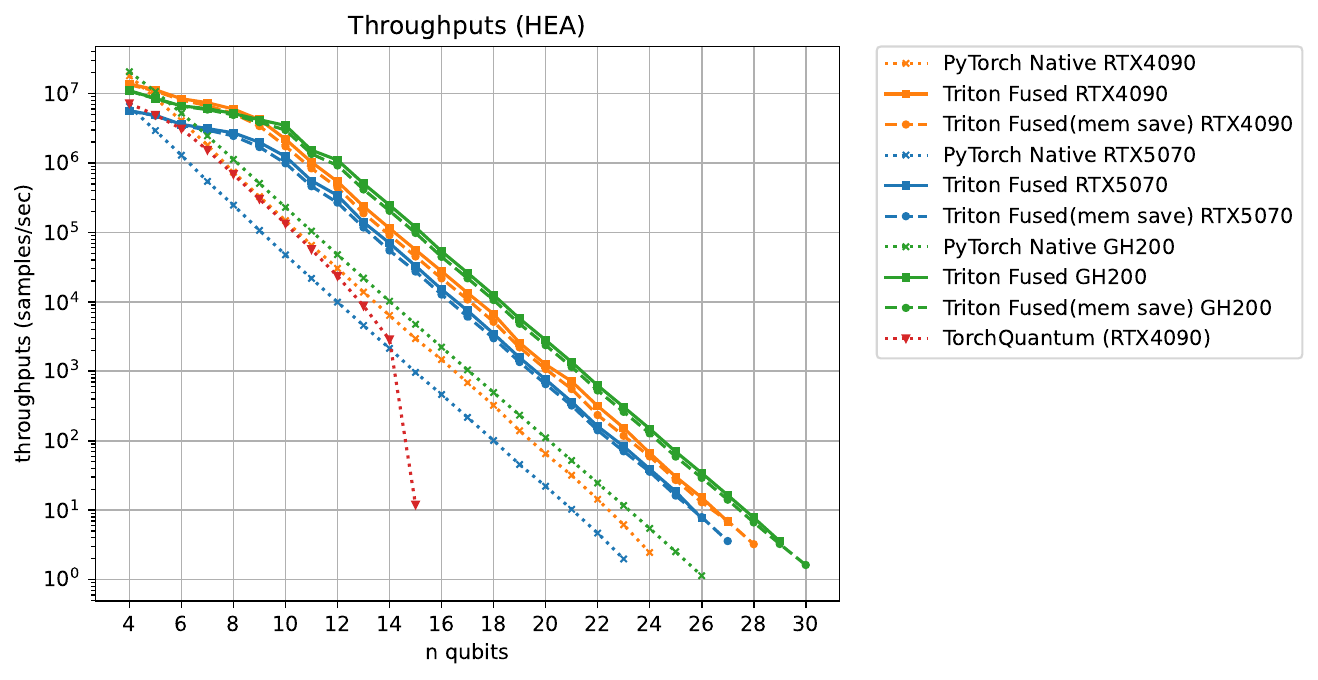}
    \caption{Throughputs for $d=1$ of the HEA shown in Fig.~\ref{fig:hea}}
    \label{fig:throughputs_xyz}
\end{figure}

Here, we measure the throughput for a task of minimizing a sum of expectation values for the observable $O=IXYZ \cdots IXYZ$ (repeating $IXYZ$ until the length matches the number of qubits) 
at the final state, made by applying an HEA (shown in Fig.~\ref{fig:hea} with $d=1$ ), consisting of {\it Rx}, {\it Ry}, {\it Rz}, and {\it CZ} gates, to random initial quantum states. 
To evaluate the throughput, we set the batch size as large as possible to be stored within the GPU memory, 
and we estimated the throughput using this batch size and the mean execution time. 
The results are shown in Fig.~\ref{fig:throughputs_xyz}. 

In this numerical experiment, to ensure a sufficient number of gates to be fused,
we fused the consecutive single-qubit gates acting on not only the same qubit but also on adjacent qubits, as shown in Fig.~\ref{fig:hea}. 
This is because fusing gates acting only on the same qubit limits the number of gates that can be fused, 
potentially making it difficult to hide memory access latency and thereby limiting the acceleration effect. 
Similarly, we also fused consecutive {\it CZ} gates, as detailed in Appendix~\ref{implementation_cnot_cz}. 

From the results presented in Fig.~\ref{fig:throughputs_xyz} and Fig.~\ref{fig:speed-up_hea}, 
we observe that for $12$ qubits and above, our proposed method achieved approximately $20$ times higher throughput compared to the PyTorch native implementation.
In particular, on the midrange GPU RTX5070 with limited memory bandwidth, our proposed method achieved approximately $30$ times higher throughput. 
These results demonstrate that implementing gate fusion in both forward and backward paths yields a significantly higher throughput than the PyTorch native implementation.  

For reference, we also plot the results of an existing QML library TorchQuantum \cite{torchquantum2024}, with a red dotted line in Fig.~\ref{fig:throughputs_xyz}. 
The implementation of this library uses a matrix-based approach (e.g., for calculating expectation values), making classical simulation for over $15$ qubits infeasible due to memory constraints. 
On the other hand, the implementation of our proposed method employs a matrix-free approach, similar to the fast quantum circuit simulator ``Qulacs" \cite{suzuki2021qulacs}. 
Furthermore, by recomputing the intermediate states of fused gates within registers during the backward path, we do not need to store them in global memory. 
As a result, for a $29$-qubit HEA on a single GPU GH200, the Triton fused implementation achieved a throughput of $3.57$ samples/sec, 
while the Triton fused (mem save) implementation achieved $3.25$ samples/sec. 
More importantly, the Triton fused (mem save) implementation successfully scaled to a $30$-qubit HEA with a throughput of $1.62$ samples/sec, 
whereas the Triton fused implementation failed at this scale due to memory constraints. 
This highlights the practical effectiveness of our memory-saving strategy for large-scale classical simulations.


\subsection{Classical simulation of $1{,}000$ layers of HEA using gradient checkpointing}\label{subsec:ckp_1000}
Finally, as a practical benchmark, 
we perform a large-scale classical simulation of a deep quantum circuit for QML 
by combining our proposed method with gradient checkpointing in PyTorch (\path{torch.utils.checkpoint}) on the GH200. 
Specifically, we perform the classical simulation of minimizing the sum of expectation values for the observable $O=IXYZ \cdots IXYZ$ (consisting of $5$ repetitions of $IXYZ$) at the final states, generated by applying a $20$-qubit HEA with $d=1{,}000$ layers (including $60{,}000$ parameters), as shown in Fig.~\ref{fig:hea}, to random initial quantum states.

\subsubsection{Preliminary experiment to determine checkpoint block size $b$}
\begin{figure}[h!]
    \centering
    \begin{subfigure}[b]{0.47\textwidth}
        \centering
        \includegraphics[width=0.95\linewidth]{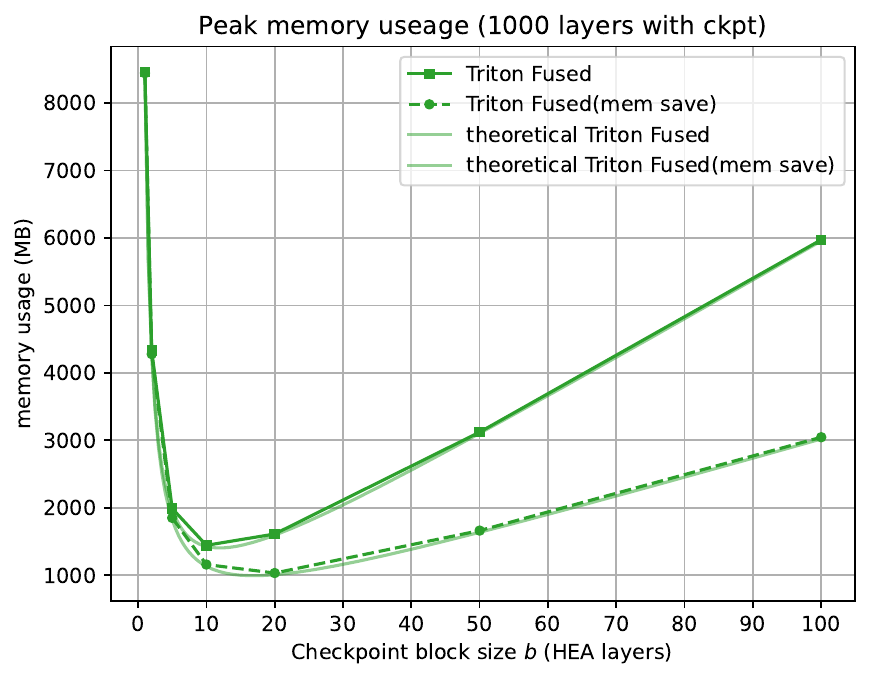}
        \caption{Peak memory usage}
        \label{fig:scale_mem}
    \end{subfigure}
    \hfill %
    \begin{subfigure}[b]{0.47\textwidth}
        \centering
        \includegraphics[width=0.95\linewidth]{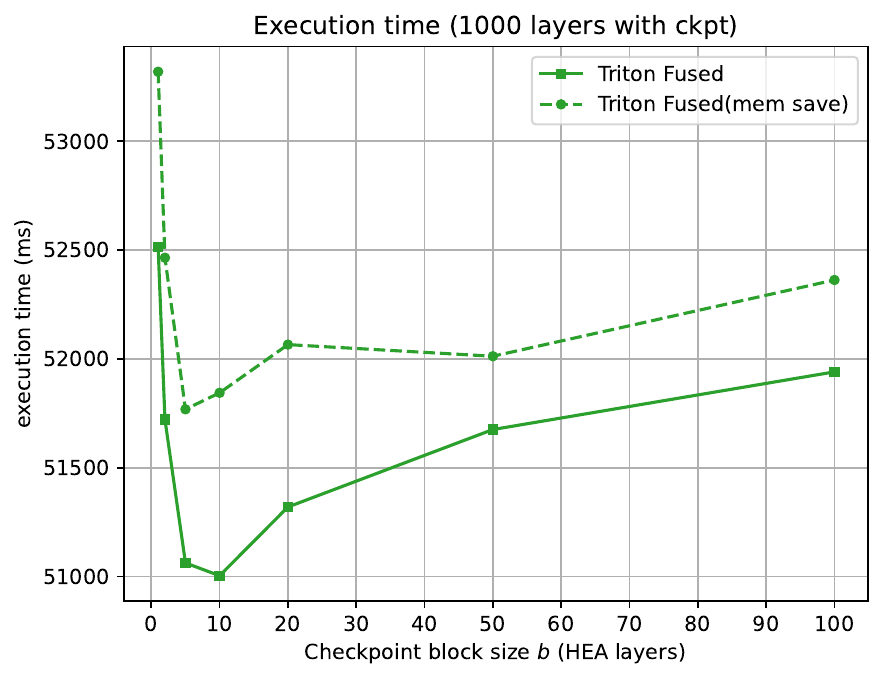}
        \caption{Execution time}
        \label{fig:scale_exec_time}
    \end{subfigure}
    \caption{ (a) Peak memory usage and (b) Execution time, 
    when varying $b$ for a $20$-qubit, $1{,}000$-layer ($d=1{,}000$) HEA with a batch size of one}
    \label{fig:scale_1}
\end{figure}

First, to optimize the checkpoint block size $b$, 
we measured the average execution time and peak memory usage with a batch size of one. 
The results are shown in Fig.~\ref{fig:scale_1}.

\textbf{Analysis of peak memory usage:}
As shown in Fig.~\ref{fig:scale_mem}, for a small $b$ ($1 \leq b \leq 5$), all implementations consume a large amount of memory. 
This is because a small block size implies a large total number of blocks ($d/b$), 
making the storage of input state vectors for each block dominant, scaling as $O(d/b)$. 
As $b$ increases, peak memory usage begins to decrease. 
Then, the minimum peak memory usage for Triton fused is $1{,}443$ MB at $b=10$, and that for Triton fused (mem save) is $1{,}032$ MB at $b=20$. 
This result demonstrates that we successfully reduced peak memory usage by approximately $30\%$ by temporarily storing the output state vectors of fused variational gates as \texttt{bfloat16}. 
As $b$ increases further, the memory required to store output state vectors for fused gates within each active block becomes dominant, causing the peak memory usage to increase again, scaling as $O(\lceil \frac{l_\text{var}}{\alpha} \rceil \cdot b)$. 
Interestingly, at $b=100$, the peak memory usage for Triton fused is $5{,}972$ MB, and that for Triton fused (mem save) is $3{,}045$ MB, which is almost half the size. 
This result indicates that when the term $\left ( \left \lceil \frac{l_\text{var}}{\alpha} \right \rceil  \cdot b \right )$ dominates, the effect of reducing peak memory usage by storing compressed state vectors is consistent with theoretical expectations.

\textbf{Comparison with theoretical analysis:}
Here, we compare the peak memory usage of our numerical results with the theoretical analysis. 
To determine $\alpha$ for the Triton fused implementation, we first focus on the number of fused gates and stored state vectors. 
As shown in Fig.~\ref{fig:hea}, for groups of three adjacent qubits, 
we fused the sequence of {\it Rx}, {\it Ry}, and {\it Rz} gates acting on each of the three adjacent qubits into a single operator ($\alpha=m=3\times 3=9$). 
For the remaining two qubits, we fused the corresponding sequence of gates into a smaller operator ($m=2 \times 3 = 6$).
So, the number of fused variational gates is $\lceil \frac{l_\text{var}}{\alpha} \rceil=\lceil \frac{60}{9} \rceil=7$, and each fused variational gate requires storing its output state vector. 
Note that a fused {\it CZ} gate does not require storing its state vectors because it does not have a variational parameter. 
Since the size of \texttt{complex64} is $8$ bytes, 
the theoretical peak memory usage for a $1{,}000$-layer HEA can be estimated by substituting $n=20$ and $d=1000$ into Eq.~\eqref{eqn:mem_cap_proposed}: 
\begin{enumerate}
    \item For the Triton fused implementation, it is given by
    \begin{equation}\label{eqn:theory_mem_numexp}
    M_{total} = (7 b + 1000/b) \times 8 \times 2^{20}. 
    \end{equation}
    \item For the Triton fused (mem save) implementation, 
    the coefficient for storing the output state vectors of fused variational gates in each active block is effectively halved because they are stored as \texttt{bfloat16} from \texttt{float32}, 
    \begin{equation*}
    M_{total} = (3.5 b + 1000/b) \times 8 \times 2^{20}. 
    \end{equation*}
\end{enumerate}
From Fig.~\ref{fig:scale_mem}, we can see that the measured values agree well with the theoretical values. 
This consistency indicates that the peak memory usage behavior of our proposed method is highly predictable, facilitating the estimation of required computational resources, even for large-scale models. 
Specifically, the theoretical optimal block sizes to minimize peak memory usage are $b^*=\sqrt{\frac{1000}{7}}\simeq 12$ for the Triton fused implementation 
and $b^*=\sqrt{\frac{1000}{3.5}}\simeq 17$ for the Triton fused (mem save) implementation. 
These values are close to our empirical observations,
where the minimum memory usage were measured at $b=10$ and $b=20$, respectively.

\textbf{Execution time and scalability:}
From Fig.~\ref{fig:scale_exec_time}, 
the execution time remains almost constant regardless of $b$ within a $1\%$ fluctuation. 
This is fundamentally because the total number of executions of fused gates remains constant regardless of $b$. 
These slight fluctuations in the execution time are caused by the overhead of calling the gradient checkpointing function, 
by reading and writing the input quantum states for the gradient checkpoints, and by the cache efficiency. 
Specifically, when $b$ is small (i.e., many checkpoints), 
the overhead of calling the gradient checkpointing function, reading and writing input quantum states for the gradient checkpoints, causes a slight increase in execution time. 
When $b$ is large, an increase in execution time may be attributed to a lower cache efficiency because a larger $b$ requires reading and writing more intermediate quantum states. 
Moreover, for Triton fused (mem save), the type conversion further increases execution time. 
Notably, the optimal block size $b$ for the fastest execution time for the Triton fused (mem save) implementation is smaller than that for Triton fused. 
This is because the input state vectors for each checkpoint block during gradient checkpointing are stored without type conversion, 
whereas the output state vectors of fused variational gates within an active checkpoint block are stored with type conversion, 
making the optimal block size $b$ for Triton fused (mem save) smaller in terms of total execution time.

\subsubsection{Numerical experiment for a practical scale QML benchmark}
Finally, assuming the training of a practical QML model using $1{,}000$ samples of randomly initialized states, 
we measured the mean training time per epoch and peak memory usage. 
For the numerical settings, we set $b=10$ for Triton fused and $b=20$ for Triton fused (mem save), based on the previous numerical results with a batch size of one. 
In addition, we set the mini-batch size to $60$ for Triton fused and $85$ for Triton fused (mem save), 
to fully use the GPU memory capacity, reaching a peak memory usage of approximately $95$ GB, as shown in Table~\ref{tbl:scale_hea}. 

As a result, the execution time was $1{,}273$ sec for Triton fused and $1{,}111$ sec for Triton fused (mem save), 
indicating that Triton fused (mem save) was approximately $10\%$ faster than Triton fused. 
Thus, the strategy of storing state vectors in reduced precision accommodates an approximately $1.4$ times larger mini-batch size, 
thereby enabling more effective utilization of GPU parallel computation resources. 
This resulted in a throughput improvement, outweighing the overhead of type conversion. 

Based on these results, for a $20$-qubit HEA model including $60{,}000$ parameters, 
we estimate the training time to be approximately $15$ to $20$ hours per epoch 
for a data set containing tens of thousands of training samples, such as MNIST or CIFAR-10. 
Moreover, since our implementation is compatible with PyTorch's standard ecosystem, 
it is easy to use multi-GPU distributed training (e.g., via DistributedDataParallel in PyTorch) to achieve even faster training. 
Therefore, our proposed method enables training deeper and larger-scale QML models within a realistic timeframe. 

\begin{table}
\caption{Performance comparison for minimizing the sum of expectation values at the final state of an HEA with $20$-qubit, $1{,}000$ layers ($60{,}000$ parameters) for $1{,}000$ samples}
\label{tbl:scale_hea}
\centering
\begin{tabular}{l| c c c c} 
\toprule
Implementation & Mini-batch size & Block size & Total execution time (sec) & Peak memory (GB)\\
\midrule
Triton Fused           & $60$ & $10$ & $1{,}273$ & $94.456$ \\
Triton Fused(mem save) & $85$ & $20$ & $1{,}111$ & $95.378$ \\
\bottomrule
\end{tabular}
\end{table}

\section{Conclusion}
In this paper, we proposed fusing quantum gates in the forward and backward paths
to improve the throughput for the classical simulation of QML and VQAs, whose performance is often limited by memory bandwidth. 
In our proposed method, instead of preserving all intermediate states, 
we store only the output state vectors of fused variational gates, 
omitting the storage of intermediate states within each fused gate and recomputing them in registers.
This approach allows us to reduce peak memory usage and communication with global memory, achieving a significant improvement in throughput. 
Specifically, we achieved an approximately $20$ times higher throughput by our proposed method, compared to the PyTorch native implementation for a typical HEA. 
In particular, on a midrange consumer GPU, the RTX 5070, we achieved an approximately $30$ times higher throughput. 
This result indicates that our proposed method is particularly effective on hardware with limited memory bandwidth. 
These results demonstrate that our approach significantly lowers the hardware barrier, enabling highly efficient classical simulations of QML and VQAs even on a single consumer GPU.

Furthermore, by combining our proposed method with gradient checkpointing, 
we demonstrated the feasibility of training a deep QML model. 
First, to optimize the checkpoint block size (i.e., the number of HEA layers to be included in the checkpoint block), 
we measured the mean execution time and peak memory usage with a batch size of one. 
The results showed that the execution time is almost constant regardless of the checkpoint block size, 
whereas the peak memory usage varies depending on the checkpoint block size, showing highly consistency with our theoretical analysis. 
In addition, we found that peak memory usage can be reduced by $30\%$ by temporarily storing the state vectors as \texttt{bfloat16} converted from \texttt{float32} (\texttt{complex64}).
Using this memory-saving implementation, 
we trained a $20$-qubit, $1{,}000$-layer HEA ($60{,}000$ parameters) on $1{,}000$ samples for one epoch, 
achieving an execution time of $1{,}111$ seconds. 
Based on this result, we estimate that we could train a $20$-qubit QML model with $60{,}000$ parameters 
on tens of thousands of training samples, such as MNIST or CIFAR-10, within $15$ to $20$ hours per epoch. 
Also, since our implementation is compatible with standard PyTorch functions, 
we can further accelerate training by distributing samples over multiple GPUs using utilities in PyTorch, such as  \texttt{DistributedDataParallel}. 
Therefore, our combined approach enables the training of deep QML models on large datasets within a realistic timeframe, which would previously require a supercomputer or a multi-GPU cluster. 

Although we focused on accelerating the training of QML models, our proposed method is expected to accelerate VQA training as well, 
since the performance of training VQAs in classical simulation is often limited by memory bandwidth. 
In addition, our implementation supports double-precision (\texttt{float64}) calculations, making it a powerful tool for theoretical research requiring higher numerical precision, such as quantum chemistry and the analysis of the barren plateau. 

We expect that our proposed method will significantly contribute to advancing the study of QML and VQAs 
by accelerating the training of wider and deeper quantum circuits with larger samples, facilitating the exploration of useful quantum applications and the verification of learning theories.

\appendix

\section*{Acknowledgements} 
This work was partially conducted using the Supermicro ARS-111GL-DNHR-LCC (Miyabi-G), equipped with GH200 Grace-Hopper Superchip, at Joint Center for Advanced High Performance Computing (JCAHPC). 

\section*{Funding} 
This work was supported by JSPS KAKENHI Grant Number JP25K21267. 

\section*{Code Availability} 
The code used in this study will be available at GitHub (\url{https://github.com/puyokw/triton_qml/tree/paper}).

\section*{Declaration of generative AI and AI-assisted technologies in the manuscript preparation process}
During the preparation of this work, we used Gemini in order to debug and refactor our code and to improve our English expressions.

\section{Analysis of the acceleration effect of classically simulating a single-layer HEA}
\begin{figure}
    \centering
    \includegraphics[width=0.9\linewidth]{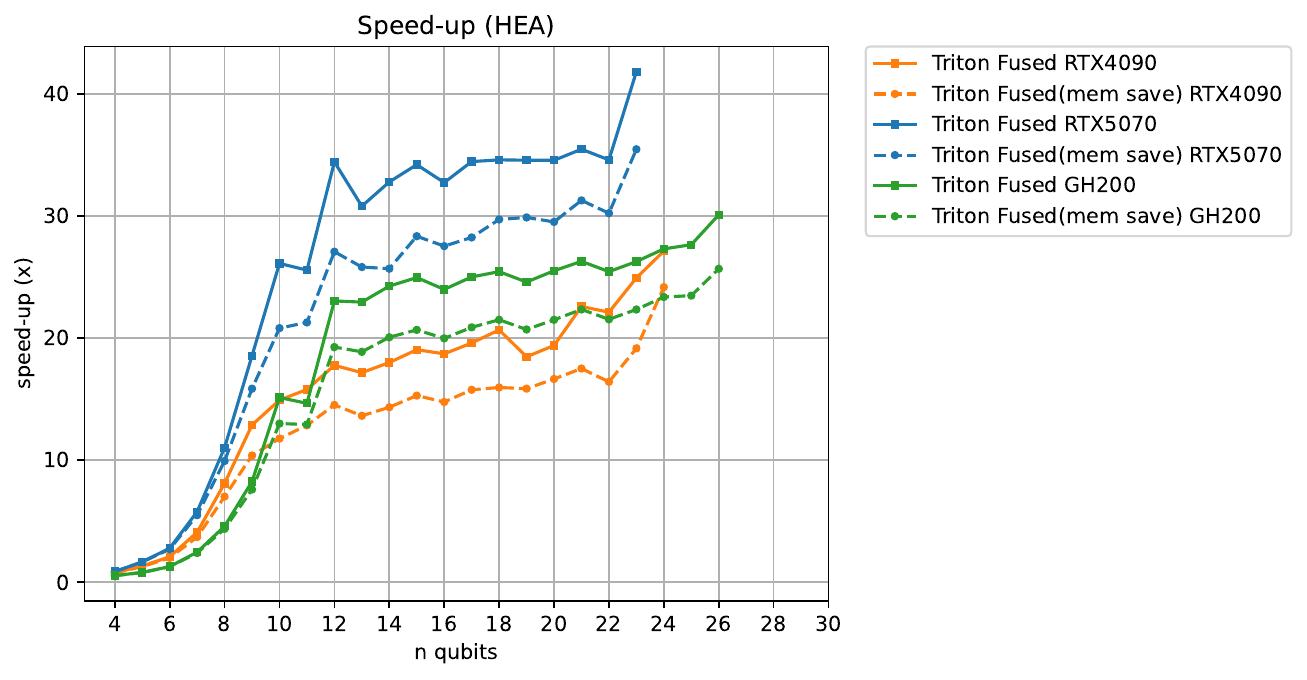}
    \caption{ Throughput improvement ratio of our proposed method relative to PyTorch native implementation, 
    based on the results in Fig.~\ref{fig:throughputs_xyz}, for a single-layer ($d=1$) of HEA in Fig.~\ref{fig:hea}. }
    \label{fig:speed-up_hea}
\end{figure}

To facilitate the performance comparison between our proposed method and the PyTorch native implementation presented in Fig.~\ref{fig:throughputs_xyz}, 
Fig.~\ref{fig:speed-up_hea} illustrates the relative throughput improvement of our proposed method compared to the PyTorch native implementation. 

From Fig.~\ref{fig:speed-up_hea}, 
we observe that the mid-range GPU RTX 5070 benefits most from our proposed method because of its relatively lower memory bandwidth, 
compared to the high-end GPU RTX 4090 and the data center GPU GH200. 
This result suggests that while the performance of the PyTorch native implementation is strongly limited by memory bandwidth, 
our proposed method effectively mitigates this bottleneck by reducing the number of accesses to global memory via gate fusion. 
We also observe that the Triton fused (mem save) implementation exhibits a slightly lower acceleration ratio than the Triton fused implementation due to the overhead of type conversion. 
Nevertheless, it remains significantly faster than the PyTorch native implementation. 

Furthermore, the periodic fluctuations are observed in Fig.~\ref{fig:speed-up_hea} (with local peaks at $n=12, 15, 18, 21, 24$), which arise from the grouping strategy of gate fusion. 
Since we fuse single-qubit gates acting on groups of three adjacent qubits, 
the computational efficiency is maximized when the number of qubits is a multiple of three.

\section{Details of other fused gate implementations}
Here, we describe the implementation of fused gates, which were not detailed in the main text. 

\subsection{The implementation of fused {\it CZ} gates and fused {\it CNOT} gates}\label{implementation_cnot_cz}
Since a {\it CNOT} gate or a {\it CZ} gate is a unitary matrix with real values ($U=U^\dagger=U^\top$) with no variational parameters, 
we can update the quantum states and adjoint states using the same kernel function for both the forward and backward paths. 
Here, we describe the implementation for the forward path. 

Let $c$ be the index of the control qubit and $t$ be the index of the target qubit. 
As we explained in Sec.~\ref{subsec:gate_basics}, 
applying a {\it CZ} gate acting on the $c$-th control qubit and the $t$-th target qubit corresponds to multiplying the probability amplitude $a_x$ by  $-1$ only when both the $c$-th and $t$-th bits of the index $x$ are $1$. 
Similarly, for the implementation of fused multiple {\it CZ} gates, 
let $c_j$ and $t_j$ denote the control and target qubits for the $j$-th gate in the fused operation. 
Each thread first counts the number of {\it CZ} gates where both the $c_j$-th and $t_j$-th bits of the index $x$ are $1$. 
If this total count is odd, the thread multiplies the corresponding probability amplitude $a_x$ by $-1$. 

Specifically, we implemented this using Triton \cite{triton2019}, as shown in Listing~\ref{code_cz}. 
First, we prepare a bitmask (\texttt{cz\_mask}) where only the $c_j$-th and $t_j$-th bits are set to $1$. 
For each thread index (\texttt{offset}), 
we check if the bitwise-AND between the mask and the index (\texttt{cz\_mask \& offset}) equals the mask itself (\texttt{cz\_mask}). 
By accumulating these boolean evaluation results across all individual {\it CZ} gates within the fused operation via an XOR operation to compute the parity, we can determine whether to flip the sign of the amplitude. 
By applying this calculation to all $2^n$ amplitudes, 
we can update the state vector for the fused {\it CZ} gates.

\begin{lstlisting}[caption=The Triton code for fused CZ gates,label=code_cz]
    parity = tl.zeros_like(offset)
    for i in tl.static_range(N_CZ_GATES):
        cz_mask = tl.load(cz_masks_ptr + i)
        parity = parity^((cz_mask & offset)==cz_mask)    
    should_flip = parity==1
    
    flipped_real = -val_real
    flipped_imag = -val_imag
    final_real = tl.where(should_flip, flipped_real, val_real)
    final_imag = tl.where(should_flip, flipped_imag, val_imag)
\end{lstlisting}

Although we used fused {\it CZ} gates in this paper, the efficient implementation of fused {\it CNOT} gates is also possible. 
Here, we describe how to implement a fused {\it CNOT} gate with out-of-place operations, 
since, as we mentioned before, PyTorch's automatic differentiation requires inputs to be invariant. 
Also, recall that applying {\it CNOT} gate corresponds to swapping the pair of amplitudes between $a_{\ldots1_c\ldots0_t\ldots}$ and $a_{\ldots1_c\ldots1_t\ldots}$ (or between $a_{\ldots0_t\ldots1_c\ldots}$ and $a_{\ldots1_t\ldots1_c\ldots}$), 
so we can implement it by tracking the index. 
Specifically, we start from the original index assigned to each thread and trace it to determine the final destination index. 
As described in Listing~\ref{code_cnot}, 
for the $j$-th {\it CNOT} gate in the fused operation, 
if the $c_j$-th bit of the current index is $1$, 
we update the current index (\texttt{out\_idx}) by applying a bitwise XOR operation between the current index and a mask (\texttt{target\_mask}) that has only the $t_j$-th bit set to $1$. 
By repeating this procedure across all individual {\it CNOT} gates within the fused operation, we obtain the final destination index. 
By writing the original probability amplitude to the resulting destination index (\texttt{out\_idx}) in a newly allocated vector, 
we can update the quantum states as an out-of-place operation.

\begin{lstlisting}[caption=The Triton code for fused CNOT gates,label=code_cnot]
    for i in range(N_GATES):
        idx = i.to(tl.int64)
        control_mask = tl.load(control_masks_ptr + idx)
        target_mask = tl.load(target_masks_ptr + idx)
        condition = (out_idx & control_mask) == control_mask
        out_idx = tl.where(condition, out_idx^target_mask, out_idx)
\end{lstlisting}

\subsection{Efficient implementation of calculating expectation values}
Here, we describe our efficient implementation for calculating the expectation values, especially for the forward path,  
because key implementation for the backward path, given by Eq.~\eqref{eqn:bwd_exp_val}, is largely shared with the forward path. 

To efficiently calculate the expectation values, 
we use a matrix-free approach without explicitly storing intermediate state vectors or operator matrices. 
As shown in Listing~\ref{code_exp_val}, first, 
to prepare the bra vector elements, 
each thread loads the probability amplitude of the final state $|\psi_M\rangle$ 
corresponding to its assigned index within the batch (\texttt{in\_batch\_idx}) and computes its complex conjugate. 

Next, we calculate the ket vector elements of $O|\psi_M\rangle$ using the relation of $Y=iXZ$. 
Since the Pauli operators $X$ and $Y$ cause bit flips, 
we determine the indexes of the probability amplitudes to be loaded using a bitwise XOR: \texttt{target\_idx=in\_batch\_idx\^{}x\_mask}, 
where \texttt{x\_mask} is a bitmask representing the target qubits for the {\it X} and {\it Y} operators in $O$. 
We then load the probability amplitude $a_{\text{target\_idx}}$, and apply the phase shifts by $Y$ and $Z$. 
In our implementation, we prepare a bitmask (\texttt{z\_mask}) representing the target qubit indexes for $Y$ and $Z$, 
and calculate the bitwise AND between \texttt{z\_mask} and \texttt{target\_idx}. 
If the parity of this result (\texttt{target\_idx}\&\texttt{z\_mask}) is odd, we multiply the amplitudes by $-1$. 
Subsequently, we multiply the amplitudes by $i=\sqrt{-1}$ to the power of \texttt{Y\_COUNT} (the total number of $Y$ in $O$). 
Note that we provide this count \texttt{Y\_COUNT} (specifically its modulo $4$, \texttt{Y\_COUNT\_MOD4}) as a compile-time constant to the Triton compiler 
in order to resolve the conditional branches during compilation, thereby avoiding the runtime overhead associated with branch divergence.

Finally, we compute the product of the obtained elements from the bra and ket vectors in each thread. 
After computing a reduction sum within each block, we calculate the global sum using \texttt{atomic\_add} operation. 
In this way, we avoid large-scale matrix multiplications and achieve efficient calculation of the expectation values 
with minimized memory accesses.

\begin{lstlisting}[caption=The Triton code for calculating expectation values,label=code_exp_val]
    in_batch_idx = offset.to(tl.int64)
    psi_ptr = state_vec_ptr + 2*in_batch_idx + 2 * transforms_per_batch * batch_idx
    psi_real = tl.load(psi_ptr, mask=mask, other=0.0)
    psi_imag = tl.load(psi_ptr + 1, mask=mask, other=0.0)
    bra_real = psi_real
    bra_imag = -psi_imag

    target_idx = in_batch_idx ^ x_mask
    ket_ptr = state_vec_ptr + 2*target_idx + 2 * transforms_per_batch * batch_idx
    ket_real = tl.load(ket_ptr, mask=mask, other=0.0)
    ket_imag = tl.load(ket_ptr + 1, mask=mask, other=0.0)

    # applying multi Z gates
    parity = _parity_jit(target_idx & z_mask)
    is_odd = parity == 1
    ket_real = tl.where(is_odd, -ket_real, ket_real)
    ket_imag = tl.where(is_odd, -ket_imag, ket_imag)
    # (multi X gates have already applied)
    if Y_COUNT_MOD4 == 0: # * 1
        pass
    elif Y_COUNT_MOD4 == 1: # * i
        # (a+bi)*i = -b+ai
        ket_real, ket_imag = -ket_imag, ket_real
    elif Y_COUNT_MOD4 == 2: # * -1
        ket_real = -ket_real
        ket_imag = -ket_imag
    elif Y_COUNT_MOD4 == 3: # * -i
        # (a+bi)*(-i) = b-ai
        ket_real, ket_imag = ket_imag, -ket_real
    
    elem_exp_val, _ = _multiply_complex_numbers(bra_real, bra_imag, ket_real, ket_imag) # real, imag
    block_exp_val = tl.sum(elem_exp_val, axis=0)
    tl.atomic_add(output_exp_val + batch_idx, block_exp_val, mask=(tl.arange(0, BLOCK_SIZE) == 0))
\end{lstlisting}

\section{Memory behavior analysis of the PyTorch native implementation using $100$-layer HEA}
In this section, to complement the experiments in Sec.~\ref{subsec:ckp_1000}, 
we describe the results of a classical simulation for $100$ HEA layers 
to verify the memory behavior of the PyTorch native implementation, which could not be evaluated in the large-scale numerical experiments in Sec.~\ref{subsec:ckp_1000} due to excessive memory requirements. 
Specifically, we verify our analytical model described in Sec.~\ref{subsec:theoretical_mem}, especially for the PyTorch native implementation, 
using a smaller model consisting of $100$ HEA layers with $20$ qubits and $6{,}000$ parameters. 
Here again, we performed the task of minimizing the sum of expectation values for the observable $O=IXYZ\ldots IXYZ$ (consisting of $5$ repetitions of $IXYZ$) at the final states, 
generated by applying the quantum circuit to random initial quantum states for a $20$-qubit system. 

Similar to Sec.~\ref{subsec:ckp_1000}, to determine the optimal size of $b$, we first measured the mean execution time and peak memory usage as we varied the checkpoint block size $b$ with a fixed batch size of $1$. 

\begin{figure}[h!]
    \centering
    \begin{subfigure}[b]{0.45\textwidth}
        \centering
        \includegraphics[width=0.9\linewidth]{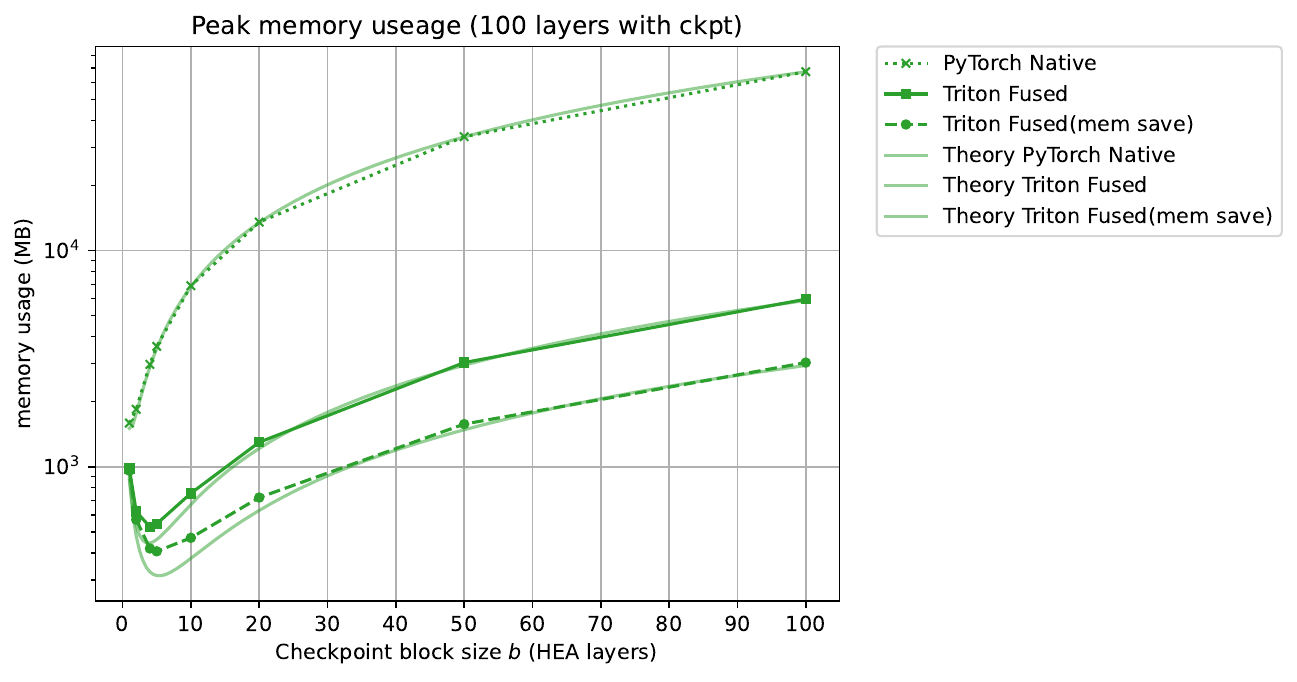}
        \caption{Peak memory usage}
        \label{fig:ckp_mem}
    \end{subfigure}
    \hfill %
    \begin{subfigure}[b]{0.45\textwidth}
        \centering
        \includegraphics[width=0.9\linewidth]{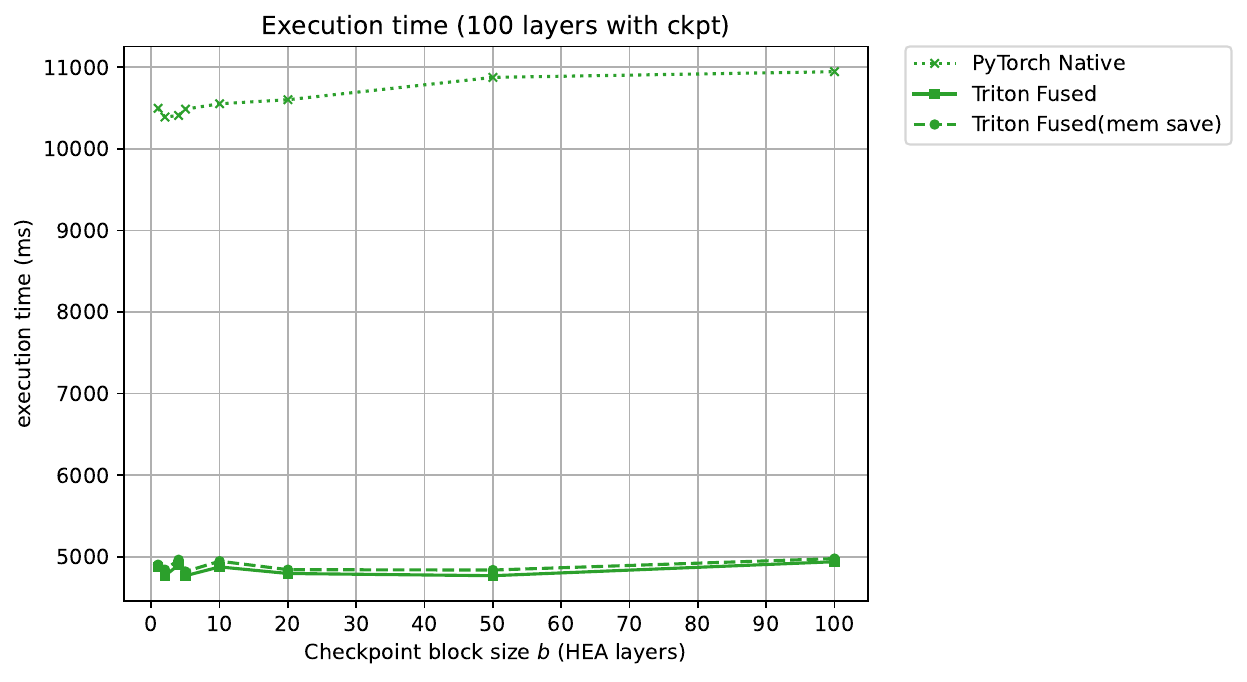}
        \caption{Execution time}
        \label{fig:ckp_exec_time}
    \end{subfigure}
    \caption{ (a) Peak memory usage and (b) Execution time, when running a $100$-layer HEA with $20$ qubits and a batch size of $1$, varying the checkpoint block size $b$}
\end{figure}

\textbf{Analyzing the behavior of peak memory usage:}
As shown in Fig.~\ref{fig:ckp_mem}, we observe that the peak memory usage for the PyTorch native implementation increases monotonically with $b$. 
This behavior is consistent with the theoretical peak memory usage. 
By substituting $l_\text{var}=60, l_\text{const}=20, d=100, M_\text{sv}=8\times 2^{20}$ (bytes) into Eq.~\eqref{eqn:mem_cap_pytorch}, 
the peak memory usage is estimated as 
\begin{equation*}
M_\text{total}\simeq \left (80 b+\frac{100}{b} \right )\times 8\times 2^{20} , 
\end{equation*}
since PyTorch's automatic differentiation stores the input quantum states for all gates, 
and each HEA layer consists of $60$ single-qubit gates and $20$ {\it CZ} gates ($80$ gates in total). 
Fig.~\ref{fig:ckp_mem} shows that the theoretical values agree well with the measured ones. 
The peak memory usage is minimized at $b^*=\sqrt{\frac{100}{80}}\simeq 1.1$. 
This analysis aligns well with the observation that peak memory usage is minimized when $b=1$. 

Similarly, the theoretical peak memory usage for the Triton fused implementation is represented by 
\begin{equation*}
M_\text{total}\simeq (7 b+\frac{100}{b})\times 8\times 2^{20} , 
\end{equation*}
which is obtained by substituting $l_\text{var}=60, \alpha=9, d=100, M_\text{sv}=8\times 2^{20}$ into Eq.~\eqref{eqn:mem_cap_proposed}, noting that $\lceil \frac{l_\text{var}}{\alpha} \rceil=7$. 
Based on this equation, the theoretical optimal block size is $b^*=3.7$, which closely matches our empirical observation where the minimum peak memory usage ($528$ MB) was recorded at $b=4$. 
For the Triton fused (mem save) implementation, the theoretical peak memory usage is estimated by 
\begin{equation*}
M_\text{total}\simeq (3.5 b+\frac{100}{b})\times 8\times 2^{20} .
\end{equation*}
So, the theoretical optimal block size for the Triton fused (mem save) implementation is $b^*=5.3$, which again aligns well with the measured minimum peak memory usage ($406$ MB) observed at $b=5$. 

Overall, as shown in Fig.~\ref{fig:ckp_mem}, we can see that the theoretical values for peak memory usage are highly consistent with the measured values across all implementations. 
Note that a slight discrepancy between theoretical and measured values is observed when memory usage is below 1 GB, primarily due to the memory overheads such as CUDA contexts. 

From Fig.~\ref{fig:ckp_exec_time}, we also see that the computation time remains almost constant regardless of the checkpoint block size $b$.

\textbf{Comparing the training time for $1$ epoch:}
Here, we compare the training times among the three implementations. 
For the checkpoint setting, based on the results of the above numerical experiments, 
we set the checkpoint block size $b$ to $b=1$ (every HEA layer) for the PyTorch native implementation,  
$b=4$ for Triton fused, and $b=5$ HEA layers for Triton fused (mem save). 
We performed the task of minimizing the sum of expectation values for the observable $O=(IXYZ)^{\otimes 5}$ at the final state, 
generated by applying $100$ HEA layers to $1{,}000$ randomly initialized quantum states with a $20$-qubit system, over $1$ epoch. 

As shown in Table~\ref{tbl:ckp_hea}, 
we set the mini-batch size to $60$ for the PyTorch native implementation, $180$ for Triton fused, and $250$ for Triton fused (mem save) so as to use approximately $90$ GB of GPU memory for each. 
The total execution times for Triton fused and Triton fused (mem save) are almost the same, and both are $18$ times faster than the PyTorch native implementation. 
This drastic acceleration is largely achieved by reducing memory access overhead and increasing the mini-batch size by three to four times through gate fusion and our memory-saving technique. 
Therefore, this numerical experiment demonstrates that our theoretical memory analysis is valid, especially for the PyTorch native implementation, 
and that our proposed method is effective for accelerating the training of QML and VQAs in medium-scale quantum circuits.

\begin{table}
\caption{ 
Results of training a $20$-qubit, $100$-layer ($6{,}000$ parameters) HEA 
for one epoch to minimize the sum of expectation values over $1{,}000$ randomly initialized states 
}
\label{tbl:ckp_hea}
\centering
\begin{tabular}{l| c c c c} 
\toprule
Implementation & Mini-batch size & Block size & Total execution time (sec) & Peak memory (GB)\\
\midrule
PyTorch Native         & $60$  & $1$ & $1{,}300$  & $89.658$ \\
Triton Fused           & $180$ & $4$ & $67.81$  & $89.993$ \\
Triton Fused(mem save) & $250$ & $5$ & $67.14$  & $91.293$ \\
\bottomrule
\end{tabular}
\end{table}

\printcredits

\bibliographystyle{model1-num-names}

\bibliography{cas-refs}


\end{document}